\pdfoutput=1
\documentclass[12pt]{article}

\usepackage{graphicx}
\usepackage{xcolor}

\usepackage[unicode,breaklinks=true,linktocpage=true]{hyperref} 
\usepackage{ascmac}

\usepackage{amsmath,amssymb,amsthm}
\usepackage{mathtools}
\numberwithin{equation}{section}

\usepackage[top=0.14\paperwidth,bottom=0.14\paperwidth,left=0.14\paperwidth,right=0.14\paperwidth]{geometry}

\usepackage{cite}

\usepackage{caption}
\DeclareMathOperator{\diag}{diag}
\DeclareMathOperator{\link}{Link}

\newcommand{\ig}{\includegraphics}

\newcommand{\vevs}[1]{\langle #1 \rangle}

\newcommand{\er}[1]{Eq.~\eqref{#1}}
\newcommand{\ers}[1]{Eqs.~\eqref{#1}}
\newcommand{\qtq}[1]{\quad\text{#1}\quad}

\newcommand{\bb}{\mathbb}
\newcommand{\na}{\nabla}
\newcommand{\hph}{\hphantom}

\newcommand{\pd}[2]{\frac{\partial{#1}}{\partial{#2}}}

\newcommand{\sr}{\sqrt}
\newcommand{\bs}{\boldsymbol}

\newcommand{\fr}{\frac}

\newcommand{\der}{\partial}

\renewcommand{\(}{\left(}
\renewcommand{\)}{\right)}

\newcommand{\dg}{\dagger}
\newcommand{\wed}{\wedge}
\newcommand{\bmx}{\left(\begin{matrix}}
\newcommand{\emx}{\end{matrix}\right)}

\begin{document}

\allowdisplaybreaks[4]
\begin{titlepage}
\hfill 
\vspace{-1em}
\renewcommand{\thefootnote}{\fnsymbol{footnote}}
\begin{flushright}
  \small RIKEN-iTHEMS-Report-24, YITP-24-145
\end{flushright}
 \begin{center}
  {\Large 
Selection rules of topological solitons 
\par 
from non-invertible symmetries
in axion electrodynamics
  \par
   }
 \vspace{1.5em} 
 
 Yoshimasa Hidaka$^{a,b}$\footnote{yoshimasa.hidaka@yukawa.kyoto-u.ac.jp}, 
 Muneto Nitta$^{c,d}$\footnote{nitta@phys-h.keio.ac.jp},
and Ryo Yokokura$^{c}$\footnote{ryokokur@keio.jp}
   \par
  \vspace{1em} 

{\small\it $^a$ Yukawa Institute for Theoretical Physics, Kyoto University, Kyoto 606-8502, Japan}

{\small\it $^b$ Interdisciplinary Theoretical and Mathematical Sciences Program (iTHEMS), RIKEN, Wako, Saitama 351-0198, Japan}

{\small\it $^c$ Department of Physics \& Research and Education Center for Natural Sciences,
\par 
Keio University, Hiyoshi 4-1-1, Yokohama, Kanagawa 223-8521, Japan}

{\small\it 
$^d$
International Institute for Sustainability with Knotted Chiral Meta Matter
(SKCM${}^{\it 2}$\!), Hiroshima University, 1-3-2 Kagamiyama, Higashi-Hiroshima, Hiroshima 739-8511, Japan
}

 \end{center}
\vspace{1.5em}
\begin{abstract}
We investigate a relation between non-invertible symmetries and selection rules of topological solitons such as axionic domain walls and magnetic strings in the $(3+1)$-dimensional  axion electrodynamics
with a massive axion or a massive photon.
In the low-energy limit of the phases where either the axion or the photon is massive, 
we identify non-invertible 0- or 1-form symmetry generators as axionic domain walls or magnetic strings, respectively.
By non-invertible transformations on magnetic monopoles or axionic strings, we give constraints on possible configurations of topological solitons in the presence of the monopoles or axionic strings.
Our results are consistent with a solution to the axionic domain wall problem by the magnetic monopole.
Further, we give a new constraint on a linked configuration of the magnetic and axionic strings.

\end{abstract}
\end{titlepage}
\setcounter{footnote}{0}
\renewcommand{\thefootnote}{$*$\arabic{footnote}} 
\tableofcontents

\section{Introduction}

Topological solitons are configurations of fields as classical solutions to equations of motion in field theories.
They have topological charges such as winding numbers, and can be extended in the spacetime.
Examples include the Abrikosov-Nielsen-Olesen magnetic vortex strings, which will be abbreviated to magnetic strings, 
in the Higgs phase of the Abelian-Higgs models as codimension-2 objects~\cite{Abrikosov:1956sx,Nielsen:1973cs} and domain walls in massive scalar theories with degenerate minima of potentials such as the sine-Gordon model as codimension-1 object~\cite{Skyrme:1958vn,Skyrme:1961vr}.
Due to the topological charges, the topological solitons can be stable under perturbations.
Topological solitons have been applied to various fields in modern physics from condensed matter, particle, and nuclear physics to cosmology (see, e.g., Refs.~\cite{Vilenkin:2000jqa,Manton:2004tk,Shnir:2005xx,Eto:2006pg,Shifman:2009zz,Weinberg:2012pjx,Eto:2013hoa,Shnir:2018yzp} for a review).

The stability of the topological solitons, however, becomes non-trivial in the presence of other objects extended along spatial or temporal directions, which may or may not be topological solitons.
One of the typical examples is the stability of the axionic domain walls in the presence of magnetic monopoles in the axion electrodynamics.
Here, the axion electrodynamics~\cite{Wilczek:1987mv} is a system with a photon and a pseudo-scalar boson called an axion~\cite{Peccei:1977hh,Weinberg:1977ma,Wilczek:1977pj,Dine:1981rt,Zhitnitsky:1980tq,Kim:1979if,Shifman:1979if} with a topological coupling between them related to a chiral anomaly.
The axionic domain walls are classical solutions to the equation of motion of the axion, which connect two different degenerate vacua of the axion potential~\cite{Sikivie:1982qv,Vilenkin:1982ks}.
The axion potential is deformed by a monopole due to the Witten effect for the axion through the topological coupling~\cite{Witten:1979ey,Fischler:1983sc}, and the stability of some domain walls can be lost~\cite{Sato:2018nqy}.
Therefore, the possible configurations of the axionic domain walls are constrained by the magnetic monopoles.
This mechanism has been applied in cosmology to a solution to the axionic domain wall problem~\cite{Sato:2018nqy}.
One may ask whether there is any underlying rule for the constraints on the topological solitons such as a selection rule, and whether it is possible to find new constraints on other topological solitons.

The purpose of this paper is 
to answer these questions in terms of a notion of generalized global symmetries~\cite{Banks:2010zn,Kapustin:2014gua,Gaiotto:2014kfa} (see also Refs.~\cite{Batista:2004sc,Pantev:2005zs,Pantev:2005wj,Pantev:2005rh, Nussinov:2006iva,Nussinov:2008pfh,Nussinov:2009zz,Nussinov:2011mz,Distler:2010zg} for earlier works, and see, e.g.,~\cite{McGreevy:2022oyu,Cordova:2022ruw,Brennan:2023mmt,Shao:2023gho,Luo:2023ive,Gomes:2023ahz} for a review). 
The generalized symmetries are defined by the existence of topological objects that can be continuously deformed.
They naturally include ordinary symmetries. Ordinary symmetries have conserved charges or symmetry generators, which can be identified as codimension-1 objects acting on 0-dimensional local operators.
The symmetry generators are topological due to conservation laws.
The conventional symmetries are invertible that are called invertible symmetries, i.e., 
the symmetry generators can have their inverses.
The generalized symmetries further include so-called higher $p$-form symmetries defined by topological objects with higher codimensions acting on $p$-dimensional objects.
The symmetry generators of generalized symmetries may or may not have their inverses, because the topological property does not necessarily require the existence of the inverse.
The symmetries without inverses are called non-invertible symmetries.
The generalized global symmetries, in particular the non-invertible symmetries, provide new selection rules~\cite{Choi:2022jqy,Cordova:2022ieu,Cordova:2022fhg,Choi:2022fgx,Chen:2022cyw,Chen:2023czk,Kaidi:2024wio}.

It has been shown that some topological solitons provide typical examples of invertible higher-form symmetries in the low-energy limit, 
in which the widths of the solitons can be neglected.
One of the virtues of identifying the topological solitons as symmetry generators in the low-energy limit is that one can find universal properties of the topological solitons which do not depend on details of models in ultra-violet theories.
For example, the magnetic string can be regarded as a 1-form symmetry generator that acts on a Wilson loop, generating 
a fractional Aharonov-Bohm phase~\cite{Banks:2010zn,Kapustin:2014gua}.
The 1-form symmetry generator 
can be characterized by the charge of the Higgs field, but does not depend on the detail of the Higgs field, e.g., the vacuum expectation value.
Similarly, the axionic domain wall can be identified as a 0-form symmetry generator that acts on the axion local operator, because the value of the axion changes when the axionic domain wall goes through the axion local operator~\cite{Kapustin:2014gua,Hidaka:2019mfm}.
The 0-form symmetry generator can be characterized by the 
so-called domain wall number, which is the 
number of the degenerate minima of the axion potential,
but does not depend on the details of the potential.
When the axion-photon coupling is turned on, the invertible symmetry generators correspond to the domain walls whose minima are not deformed by the chiral anomaly~\cite{Hidaka:2021mml,Hidaka:2021kkf}.
Meanwhile, the constraints on the domain walls, whose stability is lost by the monopoles due to the chiral anomaly, have not been explored from the viewpoint of symmetries in the low-energy limit
(see the discussions on the stability of axionic domain walls 
with the explicit breaking of the non-invertible symmetry 
by dynamical monopoles or instantons in ultra-violet models~\cite{Cordova:2023her}).

In this paper, we explore the selection rules of the topological solitons, i.e., the constraints on the possible configurations in the presence of extended objects in the $(3+1)$-dimensional axion electrodynamics from the viewpoint of the generalized symmetries, in particular, the non-invertible symmetries 
\footnote{For recent developments on generalized symmetries in the axion electrodynamics and related topics, see, e.g., Refs.~\cite{Sogabe:2019gif,Hidaka:2020iaz,Hidaka:2020ucc,Yamamoto:2020vlk,Hidaka:2020izy,Brennan:2020ehu,Heidenreich:2020pkc,Brauner:2020rtz,Heidenreich:2021xpr,Hidaka:2021mml,Hidaka:2021kkf,Yamamoto:2022vrh,Choi:2022jqy,Cordova:2022ieu,Nakajima:2022feg,Choi:2022fgx,Yokokura:2022alv,Garcia-Valdecasas:2023mis,Yamamoto:2023uzq,Cordova:2023her,Choi:2023pdp}.}.
We show that there are prohibited configurations of the topological solitons linked with or enclosing extended objects, which can be classified by 
non-invertible symmetries.

To explain the selection rules in more detail, we first focus on the axionic domain walls in the presence of the axion potential with degenerate minima, where the stability of the axionic domain walls can be violated by the magnetic monopoles.
We construct the non-invertible 0-form symmetry generators from the equation of motion for the axion using the known techniques.
We first dualize the mass term of the axion to a topological mass term with a 3-form gauge field~\cite{Aurilia:1977jz,Aurilia:1980jz,Dvali:2005an,Kaloper:2008fb}.
We then modify the topological term due to the chiral anomaly using a partition function of a 
$(2+1)$-dimensional Chern-Simons theory~\cite{Choi:2022jqy,Cordova:2022ieu} (see also Refs.~\cite{Karasik:2022kkq,GarciaEtxebarria:2022jky}).
In the low-energy limit, the non-invertible 0-form symmetry generators can be identified as axionic domain walls,
which can include the domain walls prohibited in the presence of the magnetic monopoles.
The 0-form symmetry generators can be continuously deformed by the equation of motion, but the correlation function of the 0-form symmetry generator enclosing a closed worldline of the magnetic monopole called an 't Hooft loop vanishes depending on the charge of the magnetic monopole~\cite{Choi:2022jqy,Cordova:2022ieu}.
We interpret the vanishing of the correlation function as a selection rule of the axionic domain wall that such configurations of the axionic domain walls are prohibited in the low-energy limit.

Furthermore, we find a new 
selection rule for the magnetic strings in the Higgs phase of the axion electrodynamics: the magnetic strings linked with axionic strings are prohibited depending on the charge of the axionic strings.
Here, the axionic string is a 
codimension-2 object, whose charge is the winding number of the axion around the axionic string.
The magnetic and axionic strings in the axion electrodynamics have been discussed in the context of, e.g., particle physics~\cite{Eto:2024hwn} and topological superconductors~\cite{Qi:2012cs}.

To show the selection rule,
we construct the non-invertible 1-form symmetry generator by the equation of motion for the photon based on a similar technique to that of the 0-form symmetry generator.
We again dualize the mass term of the photon to a topological mass term with a 2-form gauge field~\cite{Cremmer:1973mg,Sugamoto:1978ft,Davis:1988rw,Horowitz:1989ng,Blau:1989bq,Allen:1990gb}, and modify the topological term due to the axion-photon coupling using a partition function of a $(1+1)$-dimensional topological quantum field theory~\cite{Choi:2022fgx,Yokokura:2022alv}.
In the low-energy limit, the non-invertible 1-form symmetry generator can be regarded as a magnetic string because the generated Aharonov-Bohm phase on the Wilson loops is identical between them.
We then evaluate a correlation function of the non-invertible 1-form symmetry generator linked with the axionic string \cite{Eto:2024hwn}, which can vanish depending on the charge of the axionic string~\cite{Choi:2022fgx}.
We find that the vanishing of the correlation function can be interpreted as a selection rule that  prohibits the magnetic string 
linked with the axionic string.

In ordinary quantum theories, selection rules can be manifestly expressed using global symmetries.
In our case, we find that the symmetries that express the selection rules of the topological solitons are the invertible 1-form symmetry generators that capture the electric flux emitted from the topological solitons.

The 
selection rule for the axionic domain walls and magnetic strings can be understood from the viewpoint of ordinary quantum field theories.
The prohibited configurations of the axionic domain walls and magnetic strings have induced fractional electric charges by the Witten effect for the axion, which violates the Dirac quantization condition for the electric charge.
We show the existence of the fractional electric charge using the invertible 1-form symmetry generators, which capture the electric flux emitted from the topological solitons.

While the techniques used in this paper are based on 
the previous researches~\cite{Choi:2022jqy,Cordova:2022ieu,Choi:2022fgx,Yokokura:2022alv}, we give a new physical meaning of the non-invertible 
symmetries:
the selection rules of the topological solitons can be classified by the non-invertible symmetries.
Furthermore, one of the virtues of our results is that it is possible to apply our results to cosmological models.
Since we use a low-energy effective theory of the axion electrodynamics which does not depend on the detail of the models, our results can be universally applied to specific models.

This paper is organized as follows.
In section~\ref{sec:aed}, we introduce the action 
of the axion electrodynamics with topological masses and 
observables including topological solitons, magnetic monopoles, and axionic strings.
We then review the invertible higher-form symmetries in the presence of the topological masses in section~\ref{sec:inv}.
Section~\ref{sec:non-inv} is devoted to the construction of the non-invertible symmetry generators.
Using the non-invertible symmetry generators, we discuss the 
selection rules of the topological solitons in section~\ref{conf}.
Sec.~\ref{sec:summary}
is devoted to a summary and discussion.
In appendix~\ref{appendix}, 
we review the details of the non-invertible 0- and 1-form 
symmetry transformations on the 't Hooft loops and axionic strings,
respectively.

In this paper, we use the spacetime metric $\eta_{\mu \nu} = \diag (-1, +1, +1,+1)$ and totally anti-symmetric tensor $\epsilon_{0123} = +1$.

\section{Axion electrodynamics with topological masses\label{sec:aed}}
In this section, we review the low-energy limit of the axion electrodynamics with topological masses for the axion and photon.
After giving kinetic terms and a topological interaction, we introduce topological mass terms of the axion and photon.
We then review observables, including extended objects in the axion electrodynamics, related to the non-invertible symmetries.

\subsection{Kinetic terms and topological interaction}
We first introduce kinetic terms and an axion-photon topological coupling of the axion electrodynamics~\cite{Wilczek:1987mv},
\begin{equation}
 S_0
 = \int_{M_4} \( - \fr{v^2_\phi }{2} 
|d\phi|^2
 - \fr{1}{2e^2} |da|^2 
+ \fr{N}{8\pi^2} \phi da \wed da
\).
\label{240228.1958}
\end{equation}
Here, the axion $\phi$ is a $2\pi$-periodic pseudo-scalar field, the photon $a = a_\mu dx^\mu $ is a $U(1)$ 1-form gauge field, and $|X_p|^2$ for a $p$-form field $X_p$ is defined as $|X|^2\coloneqq X\wedge \star  X =\fr{1}{p!} X^{\mu_1 \cdots \mu_p} X_{\mu_1 \cdots \mu_p}$ with the Hodge dual operator $\star$.
The parameters $v_\phi$ and $e$ are the axion decay constant and the gauge coupling constant, respectively. 
The number $N$ is an integer, and the 4-dimensional spacetime manifold $M_4$ is assumed to be a spin manifold.
The gauge transformation laws of the axion and photon are
\begin{align}
  \phi ({\mathcal{P}})
&
\to \phi ({\mathcal{P}}) + 2\pi ,
\label{240225.2014}
\\
a & 
\to 
a +d\lambda_0,
\label{eq:gauge_trans_a}
\end{align}
respectively, where $\lambda_0$ is a $U(1)$ gauge parameter
with the large gauge transformation 
$\int_{\mathcal{C}} d\lambda_0 \in 2\pi \bb{Z}$
on a closed loop ${\mathcal{C}}$,
and ${\mathcal{P}}$ is a point in the spacetime.
Note that we regard the $2\pi$ periodic shift of 
the axion as a gauge symmetry, because it is redundancy of the description.
The fields satisfy the flux quantization condition,
\begin{equation}
 \int_{\mathcal{C}} d\phi, 
\quad
\int_{\mathcal{S}} da  \in 2\pi \bb{Z},
\label{240719.0431}
\end{equation}
where ${\mathcal{S}}$ is a 2-dimensional closed subspace. 
The quantization conditions mean that there 
can be extended objects such as 
axionic strings and magnetic monopoles.
We will discuss them in section~\ref{obs}.
The topological interaction term 
$\fr{N}{8\pi^2} \int_{M_4}\phi  da \wed da$,
which does not depend on the metric in the spacetime, 
is invariant up to $2\pi \bb{Z}$ 
under the $2\pi$ shift of the axion in \er{240225.2014}, 
because of the flux quantization condition on a spin manifold,
\begin{equation}
 \int_{M_4} da \wed da \in 8\pi^2 \bb{Z}.
\label{240719.0415}
\end{equation}

\subsection{Topological mass terms}

In the axion electrodynamics, 
the axion and photon can be massless or massive.
For example, we have a potential term 
of the axion by a non-perturbative effect, 
and a mass term of the photon by a Higgs mechanism.
The mass terms of the axion and photon can be introduced by the following actions,
\begin{equation}
 S_{\phi, \mathrm{mass}}
 = - \int_{M_4}\star V^\mathrm{axion}_{k_\phi}(\phi) ,
 \label{eq:S_mass}
\end{equation}
and 
\begin{equation}
S_{a, \mathrm{mass}} 
= -\int_{M_4} 
\(|d \xi - i k_a a \xi|^2
+ \star V^\mathrm{Higgs} (|\xi|^2) \),
\end{equation}
respectively.
The field $\xi$ is a Higgs field, which is a charge $k_a$ 
complex scalar field with the gauge transformation law $\xi \to \xi e^{i k_a \lambda_0} $.
$ V^\mathrm{axion}_{k_\phi}(\phi)$ is a potential of the axion, 
which has discrete global shift symmetry $V^\mathrm{axion}_{k_\phi}(\phi + \fr{2\pi}{k_\phi})= V^\mathrm{axion}_{k_\phi}(\phi)$.
The integer $k_\phi$ is called the domain wall number. 
We assume that $V^\mathrm{axion}_{k_\phi} (0) = (V_{k_\phi}^\mathrm{axion})' (0) =0$ and $(V^\mathrm{axion}_{k_\phi})'' (0) >0$,
but do not assume the details of the potential.
$V^\mathrm{Higgs} (|\xi|^2) $ is the Higgs potential of $\xi$, which is assumed to have minima at $|\xi| = \fr{v_a}{\sr{2}} \neq 0$
with the vacuum expectation value $\fr{v_a}{\sr{2}}$
realizing the Higgs or the superconducting phase.
The periodicity of the axion in \er{240225.2014} 
and 
the large gauge invariance of the photon in \er{eq:gauge_trans_a}
require 
that the parameters $k_\phi$ and $k_a$ should be integers, respectively.

When we discuss the low-energy limit of the axion electrodynamics, 
it will be convenient to dualize the mass terms to the topological terms.
Explicitly, the mass terms of the axion and photon can be dualized into the so-called 
$BF$-actions~\cite{Cremmer:1973mg,Sugamoto:1978ft,Davis:1988rw,Horowitz:1989ng,Blau:1989bq,Allen:1990gb,Aurilia:1977jz,Aurilia:1980jz,Dvali:2005an,Kaloper:2008fb},
\begin{equation}
 S_{\phi ,\mathrm{mass}} 
\to \fr{k_\phi}{2\pi } \int c \wed d\phi,
\end{equation}
\begin{equation}
 S_{a ,\mathrm{mass}} 
\to  \fr{k_a}{2\pi } \int b \wed da,
\end{equation}
respectively.
Here, $c$ and $b$ are 
$U(1)$ 3-, 2-form gauge fields, respectively.
The gauge transformation laws of the fields are
\begin{align}
b &
\to 
b + d\lambda_1,
\\
c 
& 
\to 
c + d\lambda_2.
\label{240229.1557}
\end{align}
Here, $\lambda_1$, and $\lambda_2$
are 1-, and 2-form $U(1)$ gauge parameters 
satisfying
\begin{equation}
\int_\mathcal{S} d\lambda_1 , \quad
\int_\mathcal{V}d \lambda_2 \in 2\pi \bb{Z},
\label{240228.2032}
\end{equation}
where the symbol $\mathcal{V}$
denotes a 3-dimensional closed space.
The gauge fields satisfy the flux quantization conditions,
\begin{equation}
\int_\mathcal{V} db, 
\quad
\int_{\Omega} dc \in 2\pi \bb{Z}.
\label{240301.0238}
 \end{equation}
Here, the symbol
$\Omega$ denotes a 4-dimensional closed space.

In the following discussion, we will consider the low-energy limit of 
the systems 
with the massive axion and the massless photon, 
or the massless axion and the massive photon.
These systems can be described by the following actions,
\begin{equation}
 S_{BF \phi }
 =  \int_{M_4} \(
 -\fr{1}{2e^2} |da|^2 
+ \fr{k_\phi}{2\pi} c \wed d \phi 
+ \fr{N}{8\pi^2} \phi da \wed da
\) ,
\label{240308.1804}
\end{equation}
and 
\begin{equation}
 S_{BF a }
 =  \int_{M_4} \( -\fr{v_\phi^2}{2} |d\phi|^2
+ \fr{k_a}{2\pi} b \wed da
+
\fr{N}{8\pi^2} \phi da \wed da
\),
\label{240718.1952}
\end{equation}
respectively.
Note that the kinetic terms of the massive fields become negligible 
in the low-energy limit.
For later convenience, we will refer to the systems 
whose low-energy limit can be 
described by the actions
$ S_{BF \phi }$ and 
$ S_{BF a }$
as 
the massive axion phase and the massive photon phase, respectively.

\subsection{Observables\label{obs}}

Here, we review 
gauge-invariant objects including 
extended objects
in the axion electrodynamics.
The extended objects can be characterized by topological charges. 
In particular, there can be 
axionic domain walls or magnetic Abrikosov-Nielsen-Olesen (ANO) vortex strings  
in the massive axion or massive photon phases, respectively.

\subsubsection{Axion local operator}

One of the basic observables in the axion electrodynamics
is the axion local operator on a point $\mathcal{P}$ in the spacetime,
\begin{equation}
    L (q_{\phi {\rm E}} , \mathcal{P})
    = e^{i q_{\phi {\rm E}}  \phi(\mathcal{P})}
\end{equation}
with a charge $q_{\phi {\rm E}} \in \bb{Z}$.
The quantization of the charge $q_{\phi {\rm E}}$
is required by the invariance under the gauge transformation 
in \er{240225.2014}.

\subsubsection{Axionic string}
The axionic string $V(q_{\phi \mathrm{M}}, \mathcal{S})$ is an object on a 2-dimensional worldsheet $\mathcal{S}$,
which has a winding number 
\begin{equation}
 \int_{\mathcal{C}} d\phi = 2\pi q_{\phi \mathrm{M}} \link (\mathcal{C,S}).
\end{equation}
Here, $\link (\mathcal{ C,S})$ is the linking number 
between ${\mathcal{C}}$ and $\mathcal{S}$.
In four dimensions, linking number between $p$- and $(3- p)$-dimensional
closed subspaces $\Sigma_p$ and $\Sigma'_{3- p}$
is defined by 
\begin{equation}
 \link (\Sigma_p, \Sigma_{3-p}')
 \coloneqq \int \delta_{p+1} (\Sigma'_{3-p}) 
\wed \delta_{3- p} (\Omega_{\Sigma_p}).
\label{240723.1651}
\end{equation}
Here, $\Omega_{\Sigma_p}$ is a $(p+1)$-dimensional subspace 
whose boundary is $\Sigma_{p}$, i.e., $\der \Omega_{\Sigma_p}= \Sigma_p$.
The symbol $\delta_{p} (\Sigma_{4-p}) $ is 
the delta function $p$-form, which is defined using a 
$(4-p)$-form field $\omega_{4 -p}$ 
as 
\begin{equation}
 \int_{M_4} \omega_{4-p} \wed \delta_p (\Sigma_{4-p})
 \coloneqq \int_{\Sigma_{4-p}} \omega_{4 -p}.
\end{equation}
Note that the linking number 
defined by \er{240723.1651} is symmetric in 
four dimensions, 
\begin{equation}
     \link (\Sigma_p, \Sigma_{3-p}')
     =  \link (\Sigma_{3-p}', 
     \Sigma_p).
\end{equation}

The existence of the axionic strings can be shown as follows.
We consider a pair of axion local operators, 
$e^{i q_{\phi {\rm E}} \phi (\mathcal{P})  -i q_{\phi {\rm E}} \phi ({\mathcal{P}'})}$
and rewrite it using the Stokes theorem 
as
\begin{equation}
    e^{i q_{\phi {\rm E}} \phi ({\mathcal{P}})  -i q_{\phi {\rm E}} \phi ({\mathcal{P}'})}
    = e^{ i q_{\phi {\rm E}} \int_{\mathcal{C_{P,P'}}} d \phi },
\end{equation}
where ${\mathcal{C_{P,P'}}}$
 is a curve whose boundaries are
 ${\mathcal{P}}$
 and ${\mathcal{P}'}$.
The choice of the curve ${\mathcal{C_{P,P'}}}$
 is not unique, 
but we can choose another curve 
${\mathcal{C'_{P,P'}}}$
with the same boundaries.
The consistency of the choices requires 
\begin{equation}
e^{i q_{\phi {\rm E}} \phi ({\mathcal{P}})  -i q_{\phi {\rm E}} \phi ({\mathcal{P}'})}
= e^{ i q_{\phi {\rm E}} \int_{\mathcal{C_{P,P'}}} d \phi }
=
e^{ i q_{\phi {\rm E}} \int_{\mathcal{C'_{P,P'}}} d \phi },
\end{equation}
i.e.,
\begin{equation}
    e^{ i q_{\phi {\rm E}} \int_{\mathcal{C}} d\phi} = 1.
\end{equation}
Here, ${\mathcal{C}} = {\mathcal{C_{P,P'}}}\cup \overline{\mathcal{C}'_\mathcal{P,P'}}$
is a closed loop, and 
 $\overline{\mathcal{C}'_\mathcal{P,P'}}$ means 
 $\mathcal{C}'_\mathcal{P,P'}$ with an opposite orientation.
Since $q_{\phi {\rm E}} \in \bb{Z}$,
we have the quantized winding number of
 the axion,
\begin{equation}
    \int_{\mathcal{C}} d\phi \in 2 \pi \bb{Z}.
\label{240314.2017}
\end{equation}
The loop ${\mathcal{C}}$ can be continuously deformed without changing the value of the 
integral, but it cannot shrink if 
$\int_{\mathcal{C}} d\phi \neq 0$,
meaning that there is a 2-dimensional object inside ${\mathcal{C}}$, which is the worldsheet of the axionic string.
We remark that in the presence of the axionic string, the axion field $\phi$
becomes a multivalued function, 
and violates the Bianchi identity, 
$dd \phi = 2\pi q_{\phi \mathrm{M}} \delta_2 ({\mathcal{S}_\phi})$.
If we try to avoid the multivalued function,
we remove a 3-dimensional small tabular neighborhood of $\mathcal{S}_\phi $
denoted by 
$\mathcal{V}_{\mathcal{S}_\phi, \epsilon}$ 
with a radius $\epsilon$
from the spacetime.
We then impose the boundary condition on 
the axion field such that it satisfies 
$\int_{\mathcal{C}_\epsilon} d \phi \in 2 \pi \bb{Z}$, where $\mathcal{C}_\epsilon \subset \der 
\mathcal{V}_{\mathcal{S}_\phi, \epsilon}$ 
is a 1-dimensional loop winding the tube 
with the radius $\epsilon$.
Since $\mathcal{V}_{\mathcal{S}_\phi, \epsilon}$ 
is removed from the spacetime, it is now clear that the loop 
of the integral ${\cal C}$ cannot shrink.

\subsubsection{\label{sec:domain-wall}Axionic domain wall}

If the axion is massive, we can have a $(2+1)$-dimensional extended object called an axionic domain wall~\cite{Sikivie:1982qv,Vilenkin:1982ks}.
The axionic domain wall is a classical solution to the equation of motion for the axion, which connects different minima of the axion potential.
For instance, the equation of motion can be 
$ v^2 d\star d \phi  - \star V'_{k_\phi}(\phi) + \fr{N}{8\pi^2} da \wed da= 0 $
if we use the action in \er{240228.1958} together with the mass term in \er{eq:S_mass}.
Since the axionic domain wall connects two different vacua, 
the configuration of the axion satisfies the condition,
\begin{equation}
 \int_{{\mathcal{P}}_\mathrm{init}}^{{\mathcal{P}}_\mathrm{fin}} \pd{\phi}{x^\mu}dx^\mu
 =
\phi ({\mathcal{P}}_\mathrm{fin}) - \phi ({\mathcal{P}}_\mathrm{init}) 
 \in \fr{2\pi   q_\mathrm{dw}}{k_\phi} \bb{Z},
\label{240228.2020}
\end{equation}
where 
$\phi ({\mathcal{P}}_\mathrm{fin}) , \phi ({\mathcal{P}}_\mathrm{init})  $ 
are the values of the axion on the vacua,
and ${\mathcal{P}}_\mathrm{fin}$ as well as ${\mathcal{P}}_\mathrm{init}$
are points at infinity.
The charge of the domain wall can be characterized by 
$q_\mathrm{dw} \in \bb{Z}$ mod $k_\phi$, which is the difference of the value
of the axion between the vacua.
Note that the charge of the domain wall $q_\mathrm{dw}$ must be 
meaningful mod $k_\phi$ because of the gauge symmetry in \er{240225.2014}.
Note that an axionic string can bound $k_\phi$ of axionic domain walls.
The equation \eqref{240314.2017} shows that there are $k_{\phi}$ of axionic domain walls around the axionic string because of $2\pi = \fr{2\pi}{k_\phi} \times k_\phi$.

In the finite energy scale, the axionic domain walls have finite widths.
Meanwhile, in the low-energy limit, the widths of the axionic domain walls can be neglected.
In this limit, an axionic domain wall can be expressed as a worldvolume of a membrane, 
\begin{equation}
 d\phi = \fr{2\pi q_\mathrm{dw}}{k_\phi} \delta_1 ({\mathcal{V}}).
\end{equation}
Here, ${\mathcal{V}}$ is a 3-dimensional subspace on which 
 the worldvolume of the axionic domain wall is defined.
Since the equation of motion for $c$ using \er{240308.1804} 
gives us $d\phi =0$ 
on the region expect for ${\mathcal{V}}$, 
the value of the axion is in the vacua,
and the relation in \er{240228.2020} can be reduced to
\begin{equation}
 \int_{\mathcal{C_{P,P'}}} d\phi 
= \int_{\mathcal{C_{P,P'}}}
\fr{2\pi  q_\mathrm{dw}}{k_\phi} \delta_1 ({\mathcal{V}}) 
= \fr{2\pi  q_\mathrm{dw}}{k_\phi} \link ((\mathcal{P,P'}),{\mathcal{V}} ),
\end{equation}
where ${\mathcal{P}}$ and ${\mathcal{P}}'$ can be points that have finite distances from the domain wall.

The axion local operator changes its value when 
it goes through the axionic domain wall.
We consider the axion local operator 
$e^{i \phi ({\mathcal{P}})}$ on a point ${\mathcal{P}}$, and deform it as $e^{i \phi ({\mathcal{P}})} \to e^{i \phi ({\mathcal{P}'})}$
such that it goes through the worldvolume of the axionic domain wall ${\mathcal{V}}$.
The deviation under the deformation is 
\begin{equation}
    e^{i \phi ({\mathcal{P}})- i \phi ({\mathcal{P}'})}
= e^{i \int_{\mathcal{C_{P,P'}}} d\phi } 
= 
e^{\fr{2 \pi i  q_\mathrm{dw}}{k_\phi} \link ({\mathcal{(P,P'), V }})}.
\label{240314.2000}
\end{equation}
Here, the integer $q_\mathrm{dw}$ denotes the number of the domain walls through which the axion local operator goes along ${\mathcal{C_{P,P'}}}$.
Therefore, $q_\mathrm{dw}$ can be identified as the charge (or the number) of the axionic domain wall.

In the above discussion, we have neglected 
the effects of the gauge field.
The stability of the domain wall becomes 
non-trivial in the presence of the electromagnetic fields.
This is due to the chiral anomaly which breaks $\bb{Z}_{k_\phi}$ 
shift symmetry explicitly.
Since the chiral anomaly 
breaks $U(1)$ symmetry to $\bb{Z}_N$ symmetry 
in the case of the massless axion, 
the remaining symmetry 
including the axion potential and the chiral anomaly 
is 
$\bb{Z}_{\gcd (k_\phi, N)}$.
Therefore, the domain walls only with $\bb{Z}_{\gcd (k_\phi, N)}$
charge are protected by the symmetry.
In particular, it has been shown in Ref.~\cite{Fischler:1983sc}
 that the potential of the axion
is deformed
in the presence of a magnetic monopole
$\bs{B} = 2\pi \cdot \fr{1}{4\pi}\fr{\bs{r}}{|r|^3}$.
Since 
the Gauss law 
$\bs{\na} \cdot \bs{E} = \fr{ e^2 N }{4\pi^2}\bs{\na} \cdot (\phi \bs{B}) $
can be solved as 
$\bs{E} = \fr{ e^2 N }{4\pi^2}\phi \bs{B}$,
the equation of motion for the axion
becomes
$ v_\phi^2 \der_\mu \der^\mu \phi - V_{k_\phi}' (\phi) =
 \fr{N}{4\pi^2} \bs{E} \cdot \bs{B}  =\fr{e^2 N^2 }{ 4 r^4}\phi $.
This means that 
the magnetic monopole induces the mass of the axion.
Therefore, 
the axionic domain wall with $\bb{Z}_{k_\phi}$ charge can be unstable
due to the mass deformation.
Note that this mechanism has been applied 
to a solution to the axionic domain wall problem in 
cosmology~\cite{Sato:2018nqy}.

\subsubsection{Wilson loop}
Next, we consider observables related to the photon.
One is a Wilson loop, which can be identified as a
closed worldline of an electric particle, 
\begin{equation}
 W (q_{a {\rm E}}, {\mathcal{C}}) = e^{iq_{a {\rm E}} \int_{\mathcal{C}} a}.
\end{equation}
Here, $q_{a {\rm E}}$ is the charge of the Wilson loop, which should be an integer,
$q_{a {\rm E}} \in \bb{Z}$ by the
invariance under the gauge transformation in \er{eq:gauge_trans_a}.
Note that this is the Dirac quantization condition 
for the electric charge.

\subsubsection{'t Hooft loop}
Another extended object is an 't Hooft loop,
which is a 1-dimensional object on a closed worldline ${\mathcal{C}}$ 
of a magnetic monopole 
with the charge $q_{a\mathrm{M}} \in \bb{Z}$.
We denote it by 
$ T (q_{q\mathrm{M}}, {\mathcal{C}})$.
Since we will not consider the dynamical excitation of the monopoles, 
the monopoles called in this paper are 't Hooft loops.
The presence of the 't Hooft loop can be characterized by the integral of $da$ 
on a closed surface,
\begin{equation}
 \int_{\mathcal{S}} da = 2\pi q_{a \mathrm{M}} \link (\mathcal{ S,C}).
\end{equation}
The existence of such an object can be shown as follows.
We can rewrite the Wilson loop 
using Stokes theorem and 
2-dimensional surfaces $\mathcal{ S_C}$ 
or $\mathcal{ S'_C}$ with the boundary 
$\der \mathcal{ S_C} = \der \mathcal{ S'_C} = {\cal C}$ as 
\begin{equation}
    e^{i q_{a {\rm E}} \int_{\mathcal{C}} a}
    = e^{i q_{a {\rm E}} \int_\mathcal{ S_C} da}
    = e^{i q_{a {\rm E}} \int_\mathcal{ S'_C} da}.
\end{equation}
The independence of the choice of surfaces 
requires 
\begin{equation}
e^{i q_{a {\rm E}} \int_{\mathcal{S}} da}
=1,
\end{equation}
where ${\mathcal{S}} =\mathcal{ S_C} \cup \bar{\mathcal{S}}'_{\mathcal{C}} $ is a 2-dimensional surface 
without boundaries.
Since $q_{a {\rm E}} \in \bb{Z}$, 
we have
\begin{equation}
    \int_{\mathcal{S}} da \in 2 \pi \bb{Z},
\end{equation}
which shows that there is a source of 
the magnetic flux inside ${\mathcal{S}}$, 
which can be identified as the magnetic monopole.
This is the Dirac quantization condition for 
the magnetic charge.
Note that in the presence of the 't Hooft loop, 
the gauge field $a$
becomes a multivalued 1-form, 
and violates the Bianchi identity, 
$d d a = 2\pi q_{a  \mathrm{M}} \delta_3 ({\mathcal{C}})$.

\subsubsection{Magnetic string}
In the low-energy limit of 
the massive photon phase, there can exist a $(1+1)$-dimensional 
extended object, which can be identified as a magnetic string.
The magnetic string is a classical solution to 
the equation of motion of the photon $a$ and the Higgs field $\xi$
in the low-energy limit of the Higgs phase.
The classical solution is called an ANO
vortex~\cite{Abrikosov:1956sx,Nielsen:1973cs},
which can be characterized by the Aharonov-Bohm phase, i.e., a line integral on a loop at infinity ${\mathcal{C}}_\infty$,
\begin{equation}
e^{ i \int_{{\mathcal{C}}_\infty} a} =e^{\fr{2 \pi  i q_\mathrm{ms}}{k_a}}  .
\end{equation}
Here, $q_\mathrm{ms} \in \bb{Z}$ mod $k_a$ is the charge of the magnetic string.
By the Stokes theorem, there is a codimension-2 
object inside 
${\mathcal{C}}_\infty$ 
where the magnetic flux is non-zero, 
\begin{equation}
    \int_{{\mathcal{C}}_\infty} a
     = \int_{{\mathcal{S}}_{{\mathcal{C}}_\infty}} da
     = \fr{2 \pi q_\mathrm{ms}}{k_a},
     \label{240301.0057}
\end{equation}
where ${\mathcal{S}}_{{\mathcal{C}}_\infty}$ is a surface whose boundary is ${\mathcal{C}}_\infty$.
While the classical solution has a finite width 
proportional to $\fr{e^2}{v_a}$, where 
the field strength $da$ is non-zero, 
the width can be neglected in the low-energy limit.
Therefore, the magnetic string can be regarded
as a worldsheet on a closed surface ${\mathcal{S}}$,
the configuration of the gauge field $a$ 
becomes
\begin{equation}
    da = \fr{2\pi q_\mathrm{ms} }{k_a} \delta_2 ({\mathcal{S}}).
\end{equation}
In the low-energy limit, the field strength vanishes 
other than the place of the magnetic strings, 
$da =0$.
The relation in \er{240301.0057} 
can also hold for any loop as
\begin{equation}
    \int_{\mathcal{C}} a 
    = \int_\mathcal{ S_C} da = \fr{2\pi q_\mathrm{ms}}{k_a}
    \link (\mathcal{ C,S}).
\end{equation}

\subsubsection{Other observables}
There are other observables in the axion electrodynamics.
Since we have 2- and 3-form gauge fields,
we can introduce the objects $e^{iq_{b {\rm E}} \int_{\mathcal{S}} b}$
and $e^{i q_{c {\rm E}}\int_{\mathcal{V}}c}$
with $q_{b {\rm E}}, q_{c {\rm E}} \in \bb{Z}$
as a generalization of the Wilson loop.
We can further introduce a 0-dimensional 
object $I(q_{b \mathrm{M}},{(\mathcal{P,P'})})$ 
with $q_{b \mathrm{M}} \in \bb{Z}$ that satisfies 
\begin{equation}
    \int_{\mathcal{V}} d b \in 2\pi
 \link (\mathcal{ V}, (\mathcal{P,P'})).
\end{equation}
This object can be understood as a generalization of
the 't Hooft loop and the axionic string.

\section{Invertible symmetries in axion electrodynamics 
with topological masses\label{sec:inv}}

In this section, we review invertible higher-form symmetries 
in the axion electrodynamics with or without 
topological masses~\cite{Hidaka:2020iaz,Hidaka:2020izy,Hidaka:2021mml,Hidaka:2021kkf}.
There are six kinds of invertible symmetries
in the low-energy limit of the massive photon or massive axion phases. 
Three of them can be obtained by equations of motion 
for the dynamical fields, 
and the others by Bianchi identities. 
We will call the symmetries associated with 
the equations of motion ``electric symmetries''
and with the Bianchi identities ``magnetic symmetries''
because they include the conservation of 
the electric and magnetic fluxes, respectively.

\subsection{Massive axion phase}

First, we consider the massive axion phase,
 where the axion is massive while the photon is massless. 
In the low-energy limit, the action is given by 
$S_{BF\phi}$ in \er{240308.1804}.

\subsubsection{Electric symmetries\label{BFphi-elec-inv}}
First, we consider the electric symmetries obtained 
from the equations of motion 
for $\phi$, 
$a$, and 
$c$,
\begin{align}
d \(\fr{k_\phi }{2\pi} c + \fr{N}{8\pi^2} a \wed da \)  &= 0 ,
\\
d \(- \fr{1}{e^2} \star da + \fr{N}{4 \pi^2} \phi da \)  &= 0 ,
\\
\fr{k_\phi }{2\pi} d\phi &= 0,
\end{align}
respectively.
Note that the equation of motion for $a$ can be understood as 
the conservation of the electric flux.
From the conserved currents, 
we can construct topological objects 
\begin{align}
U_{0, BF \phi } (e^{\fr{2\pi i m_0}{K_\phi}}, {\mathcal{V}}) 
&= \exp \(- \fr{2\pi i m_0}{K_\phi} \int_{\mathcal{V}}
\(\fr{k_\phi }{2\pi} c + \fr{N}{8\pi^2} a \wed da 
\)\), \label{eq:U0}
\\
U_{1, BF \phi} (e^{\fr{2\pi i m_1}{N}}, {\mathcal{S}}) 
&= 
\exp \(- \fr{2\pi i m_1}{N} \int_{\mathcal{S}}
\(  - \fr{1}{e^2} \star da 
+ \fr{N}{4\pi^2} \phi da
\)\),
\label{241007.0708}
\\
U_{3,  BF \phi} (e^{\fr{2\pi i m_3}{k_\phi}}, (\mathcal{ P, P'})) 
&= 
\exp \( - \fr{2\pi i m_3}{k_\phi} \cdot 
\fr{k_\phi }{2\pi} (\phi ({\mathcal{P}}) - \phi ({\mathcal{P}'}))\),
\end{align}
which can be identified as 
invertible symmetry generators,
respectively.
Here, we have introduced the following integer,
\begin{equation}
 K_\phi = \gcd (k_\phi, N),
\end{equation}
and the symmetry generators are parameterized by 
\begin{equation}
 e^{\fr{2\pi i m_0}{K_\phi}} \in \bb{Z}_{K_\phi},
\quad
 e^{\fr{2\pi i m_1}{N}} \in \bb{Z}_{N},
\quad
 e^{\fr{2\pi i m_3}{k_\phi}} \in \bb{Z}_{k_\phi}.
\end{equation}
Here, $\gcd (k_\phi, N)$ represents the greatest common divisor of $k_\phi$ and $N$.
The groups of the invertible symmetry generators cannot be continuous parameters but are restricted to the finite groups.
These restrictions come from the fact that the integrands are not gauge invariant.
For example, we consider the case of $U_{0, BF \phi } (e^{\fr{2\pi i m_0}{K_\phi}}, {\mathcal{V}})  $.
To have a gauge invariant integrand, we define the integral using 
the Stokes theorem as 
\begin{equation}
 \exp \( - i \alpha_0 \int_{\mathcal{V}}
\(\fr{k_\phi }{2\pi} c + \fr{N}{8\pi^2} a \wed da 
\)\)
\coloneqq 
 \exp \( - i \alpha_0 \int_{\Omega_{\mathcal{V}}}
\(\fr{k_\phi }{2\pi} d c + \fr{N}{8\pi^2} d a \wed da 
\)\),
\label{240310.2102}
\end{equation}
where $\Omega_{\mathcal{V}}$ is a 4-dimensional space whose boundary is 
${\mathcal{V}}$,
and $e^{i \alpha_0} \in U(1)$.
The right-hand side is manifestly gauge invariant, but we have chosen 
$\Omega_{\mathcal{V}}$ by hand.
The independence of the choices of 4-dimensional spaces requires 
\begin{equation}
  \exp \( - i \alpha_0 \int_{\Omega}
\(\fr{k_\phi }{2\pi} dc + \fr{N}{8\pi^2} d a \wed da 
\)\) =1,
\label{240723.1152}
\end{equation}
if we assume that $ U_0 (e^{i\alpha_0}, {\mathcal{V}}) $ is unitary, 
$U_0 (e^{i\alpha_0}, {\mathcal{V}})^\dg U_0 (e^{i\alpha_0}, {\mathcal{V}})  
=1 $.
Using the flux quantization conditions in \ers{240719.0415} and \eqref{240301.0238}, 
the parameter $\alpha_0$ should be 
$e^{i\alpha_0} = e^{ \fr{2\pi i  m_0}{K_\phi } } \in \bb{Z}_{K_\phi}$.

The symmetry generators act on the gauge invariant objects, 
and give us the symmetry transformation laws.
In terms of correlation functions in the path integral formalism, 
they can be written as follows
(see, e.g.,~\cite{Hidaka:2020izy,Hidaka:2021kkf} for 
the derivation)
\begin{align}
\vevs{U_{0, BF \phi} (e^{\fr{2\pi i m_0}{K_\phi} }, {\mathcal{V}}) 
e^{i q_{\phi {\rm E}} \phi ({\mathcal{P}})}
e^{- i q_{\phi {\rm E}} \phi ({\mathcal{P'}})}}
&= 
e^{  \fr{2\pi i m_0}{K_\phi} q_{\phi {\rm E}} \link ({\mathcal{V}}, ({\mathcal{P, P'}}))} 
\vevs{e^{i q_{\phi {\rm E}} \phi ({\mathcal{P}})}
e^{- i q_{\phi {\rm E}} \phi ({\mathcal{P'}})}},
\label{240301.1545}
\\
\vevs{U_{1,  BF \phi} (e^{\fr{2\pi i m_1}{N}}, {\mathcal{S}}) e^{i  q_{a {\rm E}}  \int_{\mathcal{C}} a}}
&= 
e^{ \fr{2\pi i m_1}{N} q_{a {\rm E}}  \link (\mathcal{ S,C})}
\vevs{ e^{i  q_{a {\rm E}}   \int_{\mathcal{C}} a}},
\label{240301.2157}
\\
\vevs{U_{3,  BF \phi} (e^{\fr{2\pi i m_3}{k_\phi}}, (\mathcal{ P, P'})) e^{i 
q_{c {\rm E}}  \int_{\mathcal{V}} c}}
&= 
e^{\fr{2\pi i m_3}{k_\phi} q_{c {\rm E}} \link ((\mathcal{ P, P'}),{\mathcal{V}} )}
\vevs{e^{i q_{c {\rm E}} \int_{\mathcal{V}} c}}.
\end{align}
By the dimensions of the charged objects, the 
symmetries generated by $U_0$, $U_1$, and $U_3$ 
can be identified as 0-, 1-, and 3-form symmetries, 
respectively.

Note that the restriction on the 0-form symmetry generators 
is due to the presence of the axion-photon coupling 
proportional to the integer $N$.
It is possible to understand the restriction 
for the 0-form symmetry generator 
in terms of the chiral anomaly.
Since the 0-form symmetry is the shift symmetry of the axion, 
some of the shift transformation is explicitly broken by 
the chiral anomaly, which can be characterized by the integer $N$.
Therefore, the remaining shift symmetry should be 
anomaly-free and be consistent with the periodicity of the axion potential.

We remark that the 0-form symmetry generator can be identified as 
a worldvolume of the axionic domain walls
with a charge $q_\mathrm{dw} \in \bb{Z}_{K_\phi}$,
because the axion local operator changes its value 
when the axionic domain wall goes through it
as in \er{240301.1545}.
One of the differences between the axionic domain walls in \er{240228.2020}
and $U_0$ is the charge of the domain wall.
In the case of the domain walls characterized by 
the difference of the vacua, the charge 
of the domain walls can be $\fr{2 \pi}{k_\phi} \bb{Z}$.
On the other hand, the charge of 
$U_0(e^{\fr{2\pi i m_0}{K_\phi}}, {\mathcal{V}})$ identified by 
\er{240301.1545} is restricted as 
$\fr{2\pi}{K_\phi} \bb{Z}$, and therefore 
the invertible symmetry generator 
$U_0(e^{\fr{2\pi i m_0}{K_\phi}}, {\mathcal{V}})$ only describes 
the domain walls with the restricted charges.

\subsubsection{Magnetic symmetries}

The magnetic symmetries can be obtained by the Bianchi identities,
\begin{align}
 dd \phi & =0,
\\
dd a & =0,
\\
dd c & =0.
\end{align}
We have invertible 2-, 1-, $(-1)$-form $U(1)$ 
invertible symmetries, whose 
symmetry generators are given with the real parameters 
$\beta_\phi$, $\beta_a$, and $\beta_c$ as
\begin{align}
  U_{\mathrm{2M}} (e^{i\beta_\phi }, {\mathcal{C}})
&
= 
e^{i \beta_\phi \int_{\mathcal{C}}\fr{d\phi}{2\pi} },
\\
  U_{\mathrm{1M} } (e^{i\beta_a}, {\mathcal{S}})
&
= 
e^{i \beta_a \int_{\mathcal{S}} \fr{d a}{2\pi} },
\\
  U_{\mathrm{-1M}} (e^{i\beta_c}, \Omega)
&
= 
e^{i \beta_c \int_\Omega \fr{d c}{2\pi}},
\end{align}
respectively.
Note that the normalization of the integrands are 
chosen by the flux quantization conditions 
in \er{240301.0238}.
They act on the axionic strings and 't Hooft loops as 
$U(1)$ symmetry transformations,
\begin{align}
 \vevs{U_{\mathrm{2M}} (e^{i\beta_\phi }, {\mathcal{C}}) V (q_{\phi \mathrm{M}}, {\mathcal{S}})} 
&
= 
e^{i \beta_\phi q_{\phi \mathrm{M}} \link (\mathcal{ C, S})}
\vevs{V (q_{\phi {\rm M}}, {\mathcal{S}})},
\\
 \vevs{U_{\mathrm{1M}} (e^{i\beta_a }, {\mathcal{S}}) T (q_{a \mathrm{M}}, {\mathcal{C}})} 
&
= 
e^{i \beta_a q_{a \mathrm{M}} \link (\mathcal{ S, C})}
\vevs{T (q_{a {\rm M}}, {\mathcal{C}})}.
\end{align}
Note that there is no charged object for $U_\mathrm{-1M}$.
Since the parameters satisfy $2\pi$ periodicity such as 
$\beta_\phi \to \beta_\phi + 2\pi$, these symmetry groups are 
identified as $U(1)$.

\subsection{Massive photon phase}

Using the equations of motion 
or the Bianchi identities,
the invertible symmetries in the 
low-energy limit of the massive photon phase 
can be obtained in the same way as those of the massive axion phase.
In the following, we briefly explain the invertible symmetries,
since the derivations are the same as in the case of the massive axion phase.

\subsubsection{Electric symmetries}
The equations of motions given by the action in \er{240718.1952}
are
\begin{align}
d\( v^2_\phi \star d\phi  + \fr{N}{8\pi^2} a \wed da  \) &= 0,
\\
d\(\fr{k_a }{2\pi} b 
+ \fr{N}{4\pi^2} \phi da\) & =0,
\\
\fr{k_a }{2\pi} d a &  =0.
\end{align}
Since the equations can be regarded as conservation laws, 
we have the electric symmetries with
the following symmetry generators,
\begin{align}
U_{0, BFa} (e^{\fr{2\pi i m'_0}{N}}, {\mathcal{V}}) 
&= \exp \(- \fr{2\pi i m'_0}{N} \int_{\mathcal{V}}
\(v^2_\phi \star  d\phi  + \fr{N}{8\pi^2} a \wed da 
\)\),
\label{241007.0721}
\\
U_{1, BFa} (e^{\fr{2\pi i m'_1}{K_a}}, {\mathcal{S}}) 
&= 
\exp \(- \fr{2\pi i m'_1}{K_a} \int_{\mathcal{S}}
\( \fr{k_a }{2\pi} b 
+ \fr{N}{4\pi^2} \phi da
\)\),
\label{241016.0355}
\\
U_{2, BFa} (e^{\fr{2\pi i m'_2}{k_a}}, {\mathcal{C}}) 
&= 
\exp \( - \fr{2\pi i m'_2}{k_a} \int_{\mathcal{C}}
\fr{k_a }{2\pi} a \),
\end{align}
where the parameters of the symmetry generators are
constrained due to the flux quantization conditions
in \ers{240719.0431}, \eqref{240719.0415}, and 
\eqref{240301.0238} 
by the same argument discussed in section \ref{BFphi-elec-inv} as
\begin{equation}
 e^{\fr{2\pi i m'_0}{N}} \in \bb{Z}_N,
\quad
 e^{\fr{2\pi i m'_1}{K_a}} \in \bb{Z}_{K_a},
\quad
e^{\fr{2 \pi i m'_2}{k_a}} \in \bb{Z}_{k_a}
\end{equation}
with the integer $K_a$ defined by 
\begin{equation}
 K_a = \gcd (k_a, N).
\end{equation}
Hereafter, we assume that $K_a >1$.

The symmetry transformation laws can be similarly obtained as
\begin{align}
\vevs{U_{0, BFa} (e^{\fr{2\pi i m'_0}{N}}, {\mathcal{V}}) 
e^{i   q_{\phi {\rm E}}\phi ({\mathcal{P}})}
e^{-i   q_{\phi {\rm E}} \phi ({\mathcal{P'}})}}
&= 
e^{\fr{2\pi i m'_0}{N} q_{\phi {\rm E}} \link ({\mathcal{V}}, ({\mathcal{P,P'}}))} 
\vevs{e^{i q_\phi \phi ({\mathcal{P}})}
e^{-i   q_\phi\phi ({\mathcal{P'}})}},
\label{240808.1039}
\\
\vevs{U_{1, BFa} (e^{\fr{2\pi i m'_1}{K_a}}, {\mathcal{S}}) e^{i q_{a {\rm E}}\int_{\mathcal{C}} a}}
&= 
e^{\fr{2\pi i m'_1}{K_a} q_{a{\rm E} } \link (\mathcal{ S,C})}
\vevs{ e^{i q_{a {\rm E}}  \int_{\mathcal{C}} a}},
\label{240301.2158}
\\
\vevs{U_{2, BFa} (e^{\fr{2\pi i m'_2}{k_a}}, {\mathcal{C}}) e^{i q_{b {\rm E}}  \int_{\mathcal{S}} b}}
&= 
e^{\fr{2\pi i m'_2}{k_a} q_{b {\rm E}} \link (\mathcal{ C,S})}
\vevs{e^{i q_{b {\rm E}} \int_{\mathcal{S}} b}}.
\end{align}
Note that the symmetry groups of 0- and 1-form symmetry generators 
are different from those of the massive axion phase.
The difference is due to the absence and existence of the 
mass terms of the axion and photon, respectively.

Similar to the massive axion phase, 
the 1-form symmetry generator $ U_{1, BFa}$ can be identified 
as a magnetic string with a restricted 
charge $q_\mathrm{ms} \in \bb{Z}_{K_a}$ 
because it changes the phase of the Wilson loop as in \er{240301.2158}.
The charge of the magnetic strings is, however, constrained 
by $K_a = \gcd (k_a, N)$
but not $k_a$.

\subsubsection{Magnetic symmetries}
By the Bianchi identities of the dynamical variables,
\begin{align}
 dd\phi &=0,
\\
dd a & =0,
\\ 
dd b & =0, 
\end{align}
we have the magnetic symmetries whose symmetry generators are 
\begin{align}
  U_{\mathrm{2M}} (e^{i\beta_\phi }, {\mathcal{C}})
&
= 
e^{i \beta_\phi \int_{\mathcal{C}}\fr{d\phi}{2\pi} },
\\
  U_{\mathrm{1M} } (e^{i\beta_a}, {\mathcal{S}})
&
= 
e^{i \beta_a \int_{\mathcal{S}} \fr{d a}{2\pi} },
\\
  U_{\mathrm{0M}} (e^{i\beta_b}, {\mathcal{V}})
&
= 
e^{i \beta_b \int_{\mathcal{V}} \fr{d b}{2\pi}}.
\end{align}
Note that the magnetic 1- and 2-form symmetries are the same as 
those of the massive photon phase
since the Bianchi identities are the same. 
In addition, we have a 0-form magnetic symmetry, because 
we have the 2-form gauge field $b$ in the massive photon phase.
The symmetry transformation laws are 
\begin{align}
 \vevs{U_{\mathrm{2M}} (e^{i\beta_\phi }, {\mathcal{C}}) V (q_{\phi \mathrm{M}}, {\mathcal{S}})} 
&
= 
e^{i \beta_\phi q_{\phi \mathrm{M}} \link (\mathcal{ C, S})}
\vevs{V (q_{\phi {\rm M}}, {\mathcal{S}})},
\\
 \vevs{U_{\mathrm{1M}} (e^{i\beta_a }, {\mathcal{S}}) T (q_{a \mathrm{M}}, {\mathcal{C}})} 
&
= 
e^{i \beta_b q_{a \mathrm{M}} \link (\mathcal{ S, C})}
\vevs{T (q_{a {\rm M}}, {\mathcal{C}})},
\\
 \vevs{U_{\mathrm{0M}} (e^{i\beta_b}, {\mathcal{P}}) 
I (q_{b \mathrm{M}}, {\mathcal{P}})} 
&
= 
e^{i \beta_b q_{b \mathrm{M}} \link (\mathcal{ V, P})}
\vevs{I (q_{b {\rm M}}, {\mathcal{P}})}.
\end{align}
Since the parameters of the symmetry generators are $2\pi$ periodic, 
the symmetry groups of these symmetries are $U(1)$.

\section{Non-invertible symmetries in axion electrodynamics with topological masses\label{sec:non-inv}}

In this section, we discuss the non-invertible 0- and 1-form symmetries
 in the low-energy limit of the massive axion or massive photon phases.
In the discussion of the previous section, we have invertible higher-form symmetry generators, 
which can be regarded as 
the axionic domain walls 
or the magnetic strings.
However,
there are the 
restrictions on the symmetry group due to the presence of the 
axion-photon coupling if we assume that the symmetry generators
are unitary.
Therefore, the 0- and 1-form symmetry generators 
in massive axion and photon phases cannot correspond to all of 
the axionic domain walls and the magnetic strings,
respectively.
To see the correlations 
between topological solitons 
and other extended objects 
from the viewpoint of the 
global symmetries, 
we would like to find the symmetry generators which 
can describe the topological solitons with all of the charges.

We can resolve this problem by extending these symmetry generators 
to 
the non-invertible symmetry generators 
at the expense of the unitarity.
We can construct the 0- and 1-form symmetry generators which act on the axion local operators and Wilson loops 
as $\bb{Z}_{k_\phi}$ and $\bb{Z}_{k_a}$ transformations
for the massive axion and photon phases, respectively.
While the identification of the axionic domain walls and non-invertible 0-form symmetry generators has been 
found in Refs.~\cite{Cordova:2023her},
we here explicitly construct the 0-form symmetry generators 
using the dual 3-form gauge field in the low-energy limit.

The constructions of the non-invertible
 0- and 1-form symmetry generators are essentially the same as 
those of previous researches discussed in Refs.~\cite{Choi:2022jqy,Cordova:2022ieu,Choi:2022fgx,Yokokura:2022alv}.
Meanwhile, there are two different points
from previous ones.
One is that
 the parameters of 
the non-invertible 0- and 1-form symmetry generators 
are restricted to $\bb{Z}_{k_\phi}$ and $\bb{Z}_{k_a}$
rather than arbitrary rational parameters
in the massive axion and photon phases, respectively.
The other is that the non-invertible 0- and 1-form 
symmetry generators 
can be identified as solitonic objects, i.e.,
the worldvolume of the axion domain walls and 
the worldsheet of the magnetic strings, respectively.

\subsection{Massive axion phase}

First, we consider non-invertible 0- and 1-form symmetry generators 
in the low-energy limit of the massive axion phase.
The non-invertible 0- and 1-form symmetry generators are parameterized 
by $\bb{Z}_{k_\phi}$ and $\bb{Q} / \bb{Z}$
rather than $\bb{Z}_{K_\phi}$  and $\bb{Z}_N$
in the case of the invertible symmetries, respectively.

\subsubsection{Non-invertible 0-form symmetry
and axionic domain wall}
When we discussed the constraint on the parameter $e^{i\alpha_0}$
of the invertible 0-form symmetry
in section \ref{BFphi-elec-inv},
we assumed that the symmetry generator is unitary in \er{240723.1152}.
If we relax this assumption of the invertibility, 
we can construct the symmetry generator 
parameterized by $\bb{Z}_{k_\phi}$.

The invertible 0-form symmetry generator is constructed from 
the term $e^{ - i \alpha_0 \fr{k_\phi}{2\pi} \int_{\mathcal{V}}  c}$.
and 
$e^{- i \alpha_0 \fr{N}{8\pi^2} \int_{\mathcal{V}} a \wed da}$
with a $U(1)$ parameter $e^{i\alpha_0}$ at this stage.
First, we consider the former term 
$e^{ - i \alpha_0 \fr{k_\phi}{2\pi} \int_{\mathcal{V}}  c}$.
This term cannot be modified to the best of our knowledge, 
and this term gives us the constraint 
\begin{equation}
 e^{ i \alpha_0} = e^{\fr{2\pi i n_0}{k_\phi}}\in \bb{Z}_{k_\phi}
\end{equation}
such that 
 $e^{ - i \alpha_0  \fr{k_\phi}{2\pi} \int_{\mathcal{V}} c}
 = e^{ - i n_0 \int_{\mathcal{V}} c}$
is gauge invariant.

Next, we focus on the latter term,
which is now written as 
$e^{ - i  \fr{N n_0}{4\pi k_\phi} \int_{\mathcal{V}} a \wed da}$.
This term seems to violate the gauge invariance since
the integer $k_\phi$ is in the denominator.
However, we can modify this term so that it is manifestly gauge invariant~\cite{Choi:2022jqy,Cordova:2022ieu}.
We can regard the term 
$e^{ - i \fr{ N n_0}{4\pi k_\phi} \int_{\mathcal{V}} a \wed da}$ 
as 
a naive effective action of the $U(1)$ Chern-Simon theory
whose partition function is
\begin{equation}
\begin{split}
&e^{ - i  \fr{N n_0}{4\pi k_\phi} \int_{\mathcal{V}} a \wed da}
\\
&
\to
 Z_0 [q_\phi, p_\phi ,{\mathcal{V}}, da]
 = \mathcal{ N}_0 \int \mathcal{ D}u^\phi_1... \mathcal{ D} u^\phi_{ q_\phi}
\exp\( \fr{ i }{4\pi} 
\sum_{i = 1}^{q_\phi}\int_{\mathcal{V}} 
\(
  p_\phi  u^\phi_i \wed du^\phi_i
- 2 u^\phi_i \wed da
\) \), 
\end{split}
\label{240915.1611}
\end{equation}
where 
\begin{equation}
 e^{\fr{2\pi i N n_0}{k_\phi} } = e^{ \fr{ 2\pi i q_\phi}{p_\phi}  }
\label{241008.0710}
\end{equation} 
with  co-prime integers $q_\phi, p_\phi \in \bb{Z}$,
$\mathcal{ N}_0$ is a normalization factor,
and $u^\phi_i $ ($i = 1,..., q_\phi$) are 
$U(1)$ 1-form gauge fields with 
the flux quantization condition 
$\int_{\mathcal{S}} du^\phi_i \in 2\pi \bb{Z}$.
The right-hand side is independent of the choice of 4-dimensional space
$\Omega_{\mathcal{V}}$ in the discussion around \er{240310.2102}, 
because $p_\phi$ is now in the numerator.
We thus obtain the following non-invertible 0-form symmetry generator,
\begin{equation}
 D_{0, BF \phi}
 (e^{\fr{2\pi i n_0 }{k_\phi} }, {\mathcal{V}})
= e^{ - i n_0 \int_{\mathcal{V}} c}
  Z_0 [q_\phi, p_\phi ,{\mathcal{V}},da].
\label{240314.1946}
\end{equation}
The non-invertibility comes from the fact that the 
partition function $  Z_0 [q_\phi, p_\phi ,{\mathcal{V}},da]$ 
can be zero depending on the topology of $\mathcal{V}$
and the configuration of $da$,
and therefore 
the symmetry generator cannot have 
an inverse.
We will see this in section~\ref{noninvdw}.

The symmetry transformation of the axion local operator can be 
written as
\begin{equation}
 \vevs{ D_{0, BF\phi} (e^{\fr{2\pi i n_0}{k_\phi} }, {\mathcal{V}})
 e^{i\phi ({\mathcal{P}})}
 e^{- i\phi ({\mathcal{P}})}}
=  e^{ \fr{2\pi i n_0}{k_\phi} \link ({\mathcal{V}} - {\mathcal{V}}' , ({\mathcal{P,P'}}))}
\vevs{ D_{0,BF\phi} (e^{\fr{2\pi i n_0}{k_\phi} }, {\mathcal{V}}')
 e^{i\phi ({\mathcal{P}})}
 e^{-i\phi ({\mathcal{P}})}},
\label{241021.2256}
\end{equation}
and we have the symmetry generator which generates $\bb{Z}_{k_\phi}$
transformation as desired.

By the correlation function, we can regard the 
non-invertible 0-form symmetry generator as a worldvolume 
of an axionic domain wall with $\bb{Z}_{k_\phi}$ charge.
This is because it can move continuously in the spacetime, 
and the value of the axion jumps by 
$\bb{Z}_{k_\phi}$ 
when the 0-form symmetry generator 
goes through the axion operator.

\subsubsection{Non-invertible 1-form symmetry}

Next, we consider the construction of 
the non-invertible 1-form symmetry generator.
The restriction on the symmetry group for the invertible symmetry generator
is due to the gauge invariance of 
term $\fr{N }{4\pi^2}\int_{\mathcal{S}}  \phi da $
 in \er{241007.0708}.
If the parameter is a $2\pi$ multiple of the
 rational number,
we can modify this term
by using the partition function of a 2-dimensional $BF$-theory
so that it is manifestly gauge invariant~\cite{Choi:2022fgx,Yokokura:2022alv}.
Explicitly, the modification 
for the  parameter $e^{i\alpha_1}$ with
$\alpha_1 \in 2\pi \bb{Q}$ is given by 
\begin{equation}
\begin{split}
&
 e^{ - i \alpha_1 \int_{\mathcal{S}} \fr{N}{4\pi^2}\phi da}
\\
&
\to 
Z_1 [q_a, p_a, {\mathcal{S}},da, d\phi]
= \mathcal{N}_1
\int \mathcal{ D} \chi_1^a \cdots \mathcal{ D} \chi_{ q_a}^a
\mathcal{ D} u^a_1 \cdots \mathcal{ D} u^a_{q_a}
\\
&
\qquad
\hph{
Z[q_a, p_a, {\mathcal{S}},da, d\phi]
=
\int \mathcal{ D}
}
\times
\exp \(\fr{i}{2\pi} 
\sum_{i =1}^{q_a} 
\int_{\mathcal{S}} ( p_a \chi^a_i du^a_i - 
\chi^a_i da -  u_i^a \wed d\phi) \) ,
\end{split}
\label{241008.0658}
\end{equation}
where 
$\mathcal{N}_1$ is a normalization factor,
$\chi_i^a$ $(i = 1,..., q_a)$ are $2\pi$ periodic pseudo-scalar fields, 
and $u_i^a$ are $U(1)$ 1-form gauge fields
with the normalization $\int_{\mathcal{S}} du_i^a \in 2\pi \bb{Z}$.
We have introduced co-prime integers $p_a $ and $q_a$
satisfying 
\begin{equation}
e^{ i N\alpha_1}  = e^{\fr{2 \pi i q_a}{p_a}}.
\end{equation}
In the absence of the 't Hooft loops or axionic strings, 
we can integrate the auxiliary fields, and go back to 
$ e^{ - i  \fr{q_a}{2\pi p_a} \int_{\mathcal{S}} \phi da}
= e^{ - i \alpha_1 \int_{\mathcal{S}} \fr{N}{4\pi^2} \phi da} $.
In the presence of these objects, 
the partition function becomes non-trivial, 
which we will discuss later.
Therefore, we have the following non-invertible 1-form symmetry generator, 
\begin{equation}
 D_{1, BF a} (e^{ i \alpha_1}, {\mathcal{S}})
= e^{ - i \alpha_1 \int_{\mathcal{S}} \fr{1}{e^2} \star da} 
 Z_{1, BF a} [q_a, p_a, {\mathcal{S}},da,d\phi].
\end{equation}
The symmetry generator can vanish when it acts on the 't Hooft loop or 
an axionic string, which we will see in section~\ref{axmag}.
The symmetry transformation law on the Wilson loop is 
\begin{equation}
 \vevs{ D_{1, BF a} (e^{ i \alpha_1}, {\mathcal{S}})
 e^{i\int_{\mathcal{C}} a}}
=  e^{ i\alpha_1 \link ({\mathcal{S}} - {\mathcal{S}}' , {\mathcal{C}})}
\vevs{ D_{1, BF a} (e^{i \alpha_1}, {\mathcal{S}}')
  e^{i\int_{\mathcal{C}}a} }.
\end{equation}

\subsection{Massive photon phase}
Next, we consider non-invertible 0- and 1-form symmetry generators 
in the massive photon phase.
We will show that we can construct a non-invertible 1-form symmetry 
generator that can be identified as a magnetic string with 
$\bb{Z}_{k_a}$ Aharonov-Bohm phase,
which cannot be expressed by the invertible 1-form symmetry generator
in \er{241016.0355}.
Since the constructions are similar to the case of the massive axion 
phase, we here omit the details of the constructions.

\subsubsection{Non-invertible 0-form symmetry}
Here, we consider the construction of the 
non-invertible 0-form symmetry generator.
In the massive photon phase, the restriction of the invertible 0-form symmetry generator is only due to the term 
$\fr{N}{8\pi^2}\int_{\mathcal{V}} a \wed da$ 
in \er{241007.0721}.
If the parameter is a $2\pi$ multiple of a rational number, 
we can modify this term as in the case of the massive axion phase,
and we obtain the non-invertible 0-form symmetry generator.
We introduce a parameter $\alpha'_0 \in 2\pi \bb{Q}$
with  co-prime integers $q'_\phi $ and  $p'_\phi$
satisfying 
\begin{equation}
e^{i N \alpha'_0 } = e^{ \fr{ 2\pi i q'_\phi}{p'_\phi} }.
\end{equation}
The non-invertible 0-form symmetry generator can be constructed
using the partition function in \er{240915.1611} as 
\begin{equation}
 D_{0, BF a}
 (e^{ i \alpha'_0}, {\mathcal{V}})
= e^{  -  i\alpha'_0 \int_{\mathcal{V}} v^2 \star  d\phi }
  Z_0 [q'_\phi, p'_\phi ,{\mathcal{V}},da],
\label{240316.1515}
\end{equation}
which acts on the axion local operator as
\begin{equation}
 \vevs{ D_{0, BF a} (e^{ i \alpha'_0}, {\mathcal{V}})
 e^{i\phi ({\mathcal{P}})}}
=  e^{ i \alpha'_0 
\link ({\mathcal{V}} - {\mathcal{V}}' , {\mathcal{P}})}
\vevs{ D_{0 , BF a} (e^{i \alpha'_0}, {\mathcal{V}}')
 e^{i\phi ({\mathcal{P}})}}.
\end{equation}
\subsubsection{Non-invertible 1-form symmetry
and magnetic string}

Next, we consider the construction of 
the non-invertible 1-form symmetry generator.
The symmetry generator is made of 
$e^{ - i \alpha'_1 \fr{k_a}{2\pi}\int_{\mathcal{S}}  b }$
and 
$e^{ - i \alpha'_1 \fr{N}{4\pi^2}\int_{\mathcal{S}}  \phi da }$,
where $\alpha'_1 \in 2\pi \bb{Q}$ is a parameter.
First, we focus on the former term.
The parameter $e^{i\alpha'_1}$ is constrained as 
\begin{equation}
 e^{i\alpha'_1}  = e^{ \fr{2\pi i n_1}{k_a}} \in \bb{Z}_{k_a}
\end{equation}
so that $e^{ - i \alpha'_1 \fr{k_a}{2\pi}\int_{\mathcal{S}}  b }
=e^{ - i n_1 \int_{\mathcal{S}}  b }$
is gauge invariant.
Next, we consider the latter term 
$e^{ - i \alpha'_1 \fr{N}{4\pi^2}\int_{\mathcal{S}}  \phi da }
= e^{ - i  \fr{N n_1}{2\pi k_a }\int_{\mathcal{S}}  \phi da }$,
which seems to violate the gauge invariance, again.
We can modify this term
by using the partition function of a 2-dimensional $ BF$ theory
in \er{241008.0658} and obtain the non-invertible 1-form symmetry 
generator,
\begin{equation}
 D_{1, BFa } (e^{\fr{2\pi i n_1}{k_a} }, {\mathcal{S}})
= e^{ - i n_1 \int_{\mathcal{S}} b} 
 Z_1 [q'_a, p'_a, {\mathcal{S}},da , d\phi],
\end{equation}
where we have introduced co-prime integers $q'_a$ and $p'_a$
satisfying 
\begin{equation}
 e^{\fr{2\pi i N n_1}{k_a}} = e^{\fr{ 2\pi i q'_a}{p'_a}}.
\end{equation}

The non-invertible 1-form symmetry generator can be identified as 
the worldsheet of the magnetic string with $\bb{Z}_{k_a}$
charge,
because it can move continuously and can act on the Wilson loop and give 
a $\bb{Z}_{k_a} $-valued 
fractional Aharonov-Bohm phase,\begin{equation}
 \vevs{ D_{1, BFa} (e^{\fr{2\pi i n_1}{k_a} }, {\mathcal{S}})
 e^{i\int_{\mathcal{C}} a}}
=  e^{ \fr{ 2\pi i n_1}{k_a} \link ({\mathcal{S}} - {\mathcal{S}}' , {\mathcal{C}})}
\vevs{ D_{1, BFa }(e^{\fr{2\pi i n_1}{k_a} }, {\mathcal{S}}')
  e^{i\int_{\mathcal{C}}a} }.
\end{equation}
We remark that the possible 
phase of the Aharonov-Bohm phase is the same as the magnetic strings.
This is in contrast to the invertible symmetry generators,
where the possible Aharonov-Bohm phase is constrained as 
$\bb{Z}_{K_a}$.

\section{Selection rules 
of topological solitons from non-invertible symmetries\label{conf}}

In this section, 
we show that the non-invertible symmetries 
lead to selection rules for 
the topological solitons
in the presence of magnetic monopoles or axionic strings.
In the previous section, we show the existence of 
non-invertible symmetries of electric 0- and 1-form symmetries.
The symmetry generators 
can be identified as the axionic domain walls and 
magnetic strings depending on the phases, respectively.

One may think that the topological stability of the symmetry generators
would contradict the violation of the stability 
of the axionic domain walls 
due to the monopoles
explained in section~\ref{sec:domain-wall}.
The apparent contradiction can be resolved by 
the non-invertible 0-form symmetry transformation on the monopole,
which can be expressed
by the vanishing of the  correlation function of the non-invertible 0-form symmetry generator and the 't Hooft loop~\cite{Choi:2022jqy,Cordova:2022ieu}.

We can further find the selection rule for the magnetic string 
in the massive photon phase:
the magnetic string linked with the axionic string is prohibited
depending on the charge of the magnetic and axionic strings.
We can derive the selection rule using the non-invertible 
1-form transformation on the axionic string~\cite{Choi:2022fgx}.

As conventional selection rules can be expressed in terms of
global symmetries,
we find that 
our selection rules can be expressed in terms of the invertible electric 
1-form symmetries.
By this discussion,
we give an interpretation 
of the selection rules of the solitons 
in the language of the conventional quantum field theories.
In both cases, the prohibited configurations violate the Dirac quantization condition.
While we derive the selection rules in the low-energy limit,
we argue that we can apply our selection rules to the energy region where the widths of the solitons are finite.

\subsection{\label{noninvdw}
Selection rule 
for axionic domain wall
linked with monopole}

First, we consider the 
non-invertible 0-form symmetry transformation on 
the 't Hooft loop.
As shown in Refs.~\cite{Choi:2022jqy,Cordova:2022ieu},
the correlation function of 
the 0-form symmetry generator 
linked with the 't Hooft loop 
vanishes.
By the correlation function,
we find the selection rule of
 the axionic domain wall.
 We relate the selection rule to the Dirac quantization condition using the invertible 1-form symmetry.

\subsubsection{Non-invertible 0-form symmetry 
transformation on 't Hooft loop}
\begin{figure}[t]
 \begin{center}
  \ig[height=10em]{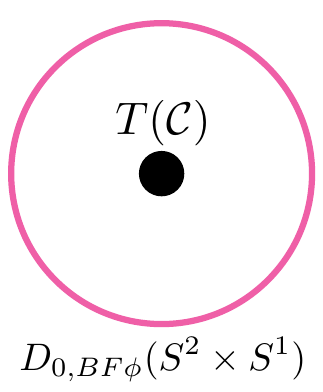}
 \end{center}
\caption{\label{dw-mon}The configuration of the 't Hooft loop 
and non-invertible 0-form symmetry generator.
This figure is a spatial intersection of the time slice of the configuration.
The Black bullet and pink circle represent 
the time slice of the 't Hooft loop
and the non-invertible 0-form symmetry generator, respectively.
We omit the charge or the parameter of the objects to simplify the notation.
On the time slice, there should be another bullet of 
the 't Hooft loop linked with the non-invertible 0-form symmetry generator
with opposite charges, but we omit them.
}
\end{figure}

Here, we consider the configuration where 
the 0-form symmetry generator encloses an 't Hooft loop.
For concreteness, we consider the 
't Hooft loop on 
a temporally extended 
loop ${\cal C}$ with the topology of $S^1$.
We take
the 3-dimensional subspace 
${\cal V}$
of the 
non-invertible 0-form symmetry generator
in \er{240314.1946}
as $ {\cal V} = S^2 \times S^1$
with the spatially 
extended $S^2$
and the temporally extended $S^1$.
On each time slice, 
the non-invertible 
symmetry generator 
encloses the 't Hooft loop without intersections
(see Fig.~\ref{dw-mon}).
Note that this configuration with $k_\phi =1$ 
has been considered in Refs.~\cite{Kogan:1992bq,Kogan:1993yw},
and is called a monopole bag.
Hereafter, we will refer to configurations 
of extended objects that are 
not intersected but topologically non-trivial
as linked configurations.

In this configuration, the correlation function 
between the non-invertible 
0-form symmetry generator 
$D_{0, BF \phi }(e^{\fr{2 \pi i n_0 }{ k_\phi}} ,S^2 \times S^1)$
with the parameter 
$\fr{N n_0}{k_\phi} = \fr{q_\phi}{p_\phi}$
in \er{241008.0710}
and the 't Hooft loop 
$T(q_{a\mathrm{M}}, {\cal C}) $
can vanish~\cite{Choi:2022jqy,Cordova:2022ieu} 
(see appendix \ref{0noninv} for a review of the derivation),
\begin{equation}
    \vevs{
    D_{0, BF \phi }(e^{\fr{2 \pi i n_0 }{ k_\phi}} ,S^2 \times S^1)
    T(q_{a\mathrm{M}}, {\cal C}) 
    } =  \delta_{q_{a \mathrm{M}}, 0}^{\mathrm{mod}\, p_\phi} 
\vevs{T(q_{a\mathrm{M}}, {\cal C}) 
e^{ -i\fr{q_\phi q_{a \mathrm{M}}}{p_\phi} \int_{S^1} a_\mathrm{R}}},
\label{240917.1539}
\end{equation}
where
\begin{equation}
    \delta_{0,  q_{a\mathrm{M}}
}^{\mathrm{mod}\,  p_\phi} 
= 
\begin{cases}
1,  & q_{a \mathrm{M}}
 = 0 \mod p_\phi,
 \\
 0, & \text{otherwise}.
\end{cases}
\end{equation}
Here, $a_\mathrm{R}$ represents the regular part of the photon
under the decomposition to the regular and singular parts,
\begin{equation}
    a = a_\mathrm{R} + a_\mathrm{S},
\label{241016.0441}
\end{equation}
with $dda_{\rm R} =0$ but
$d da_{\mathrm{S}} \neq 0$ in the presence of the monopole.
The induced Wilson loop 
$e^{ -i\fr{q_\phi q_{a \mathrm{M}}}{p_\phi} \int_{S^1} a_\mathrm{R}}$ 
with the charge $-\fr{q_\phi q_{a \mathrm{M}}}{p_\phi} $
is required by the electric charge conservation,
since we have the induced charge on the axionic domain wall
due to the Witten effect for the axion.

\subsubsection{Selection rule for axionic domain wall
and Dirac quantization condition}
The vanishing of the correlation function 
in quantum theories
implies that the 
vanishing of the 
transition amplitude.
In the low-energy limit of the quantum field theories, 
the vanishing of the transition amplitude implies 
some physical meanings such as 
the confinement of charged particles and the 
selection rules.

We argue that 
the vanishing of the correlation function can be identified as a selection rule.
This means that the axionic domain wall linked with the magnetic monopole is prohibited depending on the charge of the magnetic monopole and the axionic domain wall.
We can characterize 
the selection rule by using a global symmetry
as in ordinary selection rules.
Here we use the invertible electric 1-form global symmetry.

\begin{figure}[t]
 \ig[width=35em]{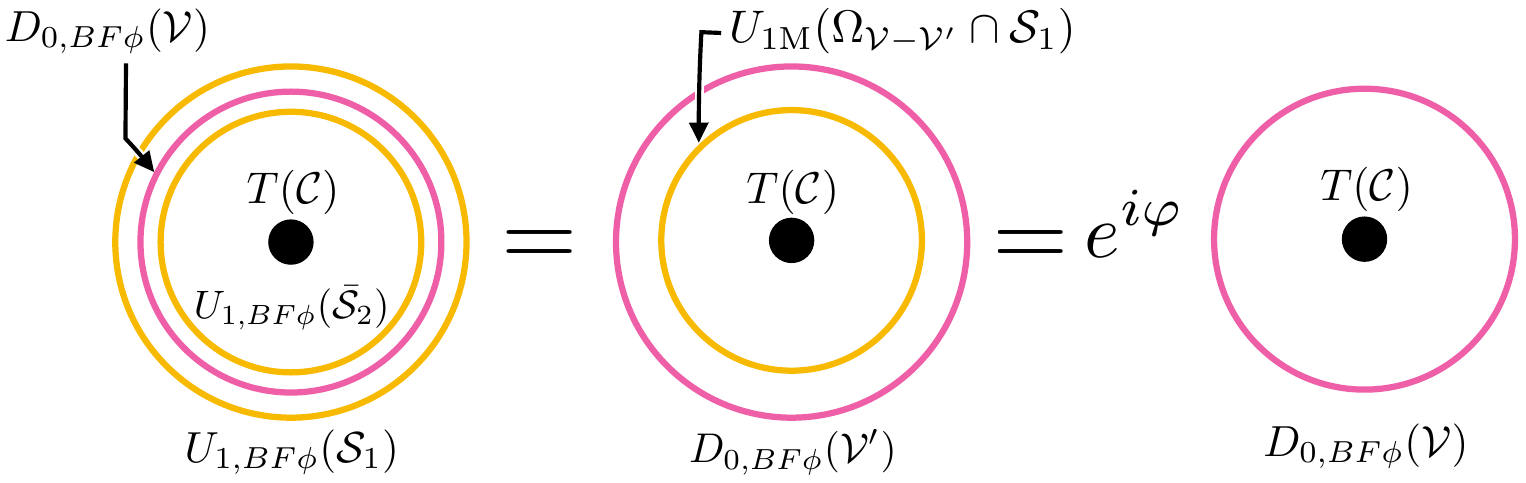}
\caption{\label{inv1-noninv0}
Correlation function of an 't Hooft loop, 
a non-invertible 0-form symmetry generator, and invertible electric 1-form symmetry generators in \ers{240716.0425}--\eqref{240716.0426}.
In addition to the objects in Fig.~\ref{dw-mon},
the yellow circles in the left panel 
represent the intersections of the 
instantaneous spheres of the invertible electric 1-form symmetry generators 
in \er{240716.0425}.
In the middle panel, 
magnetic 1-form symmetry generator represented by 
the yellow circle is induced under the continuous deformation 
${\cal V} \to {\cal V'}$
of the non-invertible 0-form symmetry generator
in \er{241022.0400}.
In the right panel, the magnetic 1-form symmetry generator 
acts on the 't Hooft loop and gives the phase 
$e^{i\varphi} = e^{ - \fr{2\pi i m_1}{N} \cdot \fr{q_\phi}{p_\phi} \cdot q_{a \mathrm{M}}}$ in \er{240716.0426}.
}
\end{figure}

This violation can be seen 
by capturing the electric 
flux emitted from 
the axionic domain wall 
by the invertible electric 1-form symmetry generator.
We consider the following correlation function of 
the 't Hooft loop on ${\mathcal{C}}$, the non-invertible 0-form symmetry 
generator on ${\mathcal{V}}$, 
and the invertible electric 1-form symmetry generator
on ${\mathcal{S}}_1$ and $\bar{\mathcal{S}}_2$,
\begin{equation}
 \vevs{U_{1, BF \phi} 
(e^{\fr{2\pi i m_1}{N}}, {\mathcal{S}}_1 \cup \bar{\mathcal{S}}_2)
D_{0, BF \phi} (e^{\fr{2\pi i n_0 }{k_\phi}}, {\mathcal{V}}) T(q_{a \mathrm{M}}, {\mathcal{C}})},
\label{240716.0425}
\end{equation}
where ${\mathcal{S}}_1$ and $\bar{\mathcal{S}}_2$ enclose the 
non-invertible 0-form symmetry generator at a time slice
(Fig.~\ref{inv1-noninv0}).
Physically, the invertible electric 1-form symmetry generator captures 
the electric flux from the induced electric charge on 
the 0-form symmetry generator.
We then move the 0-form symmetry generator ${\mathcal{V}} \to \mathcal{ V'}$ 
so that 
${\mathcal{S}}_2$ and ${\mathcal{S}}_1$ are pair-annihilated.
By this deformation, the non-invertible 0-form symmetry generator 
acts on the invertible 1-form symmetry generator 
as an invertible translation 
$\phi \to \phi + \fr{2 \pi n_0 }{k_\phi} \delta_0 
(\Omega_{{\mathcal{V}} - \mathcal{ V'}}) $,
where $\Omega_{{\mathcal{V}}  - \mathcal{ V'}}$ is a 4-dimensional space 
with $\der \Omega_{\mathcal{ V} - {\mathcal{V'}}} = 
\mathcal{ V} \cup \bar{\mathcal{V}}'$.
By the action, the invertible magnetic 1-form symmetry generator is 
induced, 
\begin{equation}
\begin{split}
&
  \vevs{U_{1, BF \phi} 
(e^{\fr{2\pi i m_1}{N}}, {\mathcal{S}}_1 \cup \bar{\mathcal{S}}_2)
D_{0, BF \phi} (e^{\fr{2\pi i n_0 }{k_\phi}}, {\mathcal{V}}) T(q_{a \mathrm{M}}, {\mathcal{C}})}
\\
&
= 
  \vevs{U_\mathrm{1 M} 
(e^{ - \fr{2\pi i m_1}{N} \cdot \fr{q_\phi}{p_\phi}},
{\mathcal{S}}_1)
D_{0, BF \phi} (e^{\fr{2\pi i n_0 }{k_\phi}}, {\mathcal{V}}') T(q_{a \mathrm{M}}, {\mathcal{C}})} 
\end{split}
\label{241022.0400}
\end{equation}
Here, we have used 
$\Omega_{\mathcal{ V} - {\mathcal{V}}'} \cap {\mathcal{S}}_1 = {\mathcal{S}}_1$
by the choice of $\Omega_{\mathcal{ V} - {\mathcal{V}}'} $.
Since the magnetic 1-form symmetry generator only acts on the 't Hooft loop 
in the correlation function, 
we arrive at the relation
after moving $\mathcal{ V'}$ to ${\mathcal{V}}$ again,
\begin{equation}
\begin{split}
&
  \vevs{U_{1, BF \phi} 
(e^{\fr{2\pi i m_1}{N}}, {\mathcal{S}}_1 \cup \bar{\mathcal{S}}_2)
D_{0, BF \phi} (e^{\fr{2\pi i n_0 }{k_\phi}}, {\mathcal{V}}) T(q_{a \mathrm{M}}, {\mathcal{C}})}
\\
&
= 
e^{ - \fr{2\pi i m_1}{N} \cdot \fr{q_\phi}{p_\phi} \cdot q_{a \mathrm{M}}}
  \vevs{
D_{0, BF \phi} (e^{\fr{2\pi i n_0 }{k_\phi}}, {\mathcal{V}}) T(q_{a \mathrm{M}}, {\mathcal{C}})} .
\end{split}
\label{240716.0426}
\end{equation}
However, we have a problem with the correlation function.
Since the invertible electric 1-form symmetry generator is parameterized by 
$e^{\fr{2\pi i m_1}{N}} \in \bb{Z}_N$, 
the correlation function should be invariant under
$m_1 \to m_1 + N$ due to the periodicity of $ \bb{Z}_N$.
The invariance requires that 
$e^{-  \fr{2\pi i  N n_0}{k_\phi} \cdot q_{a \mathrm{M}}} = 
e^{-  \fr{2\pi i q_\phi q_{a \mathrm{M}}}{p_\phi} } =1$,
where we have used $\fr{N n_0}{k_\phi} = \fr{q_\phi}{p_\phi}$
defined in \er{241008.0710}.
This is only satisfied for $q_{a\mathrm{M}} \in p_\phi \bb{Z}$.
Otherwise, the correlation function becomes ambiguous,
i.e., the $\bb{Z}_{p_\phi}$ phase cannot be determined.
To avoid this ambiguity, the correlation function
$\vevs{
D_{0, BF \phi} (e^{\fr{2\pi i n_0 }{k_\phi}}, {\mathcal{V}}) T(q_{a \mathrm{M}}, {\mathcal{C}})} $ should vanish,
\begin{equation}
 \vevs{
D_{0, BF \phi} (e^{\fr{2\pi i n_0 }{k_\phi}}, {\mathcal{V}}) T(q_{a \mathrm{M}}, {\mathcal{C}})}  = 0 ,\qtq{for} q_{a \mathrm{M}} \not\in  p_\phi \bb{Z},
\end{equation}
which is the same result as 
the explicit evaluation of the partition function in \er{240917.1539}.

In the language of ordinary quantum field theories, 
the phase ambiguity in the 1-form symmetry transformation can be understood as the violation of the 
Dirac quantization condition.
By the equation of motion for the photon,
$\fr{1}{e^2} d\star da =   \fr{N}{4\pi^2 } d\phi \wed da$,
we have the induced electric charge on 
a 3-dimensional subspace ${\cal V_S}$ with 
the boundary ${\cal S}$,
$\fr{N}{4\pi^2 } \int_\mathcal{ V_S} d\phi \wed da 
= \fr{N}{4\pi^2 } \int_\mathcal{ V_S} \bs{\na} \phi \cdot \bs{B} d V $.
By taking ${\mathcal{S}}$ as a spatially extended 2-dimensional subspaces 
which enclose the axionic domain wall,
we find 
$\fr{N}{4\pi^2 } \int_\mathcal{ V_S} \bs{\na} \phi \cdot \bs{B} d V
 =  - \fr{ N  n_0 q_{a\mathrm{M}}}{k_\phi } 
= - \fr{ q_\phi q_{a\mathrm{M}}}{p_\phi } $ 
mod 
$\bb{Z}$.\footnote{Note that $\bs{\na} \phi \cdot \bs{B} < 0 $ 
in our choice so that the phase in \er{241021.2256} is positive.} 
If the 't Hooft loop with the charge $q_{a{\rm M}}$
is  $q_{a{\rm M}} \not\in  p_\phi \bb{Z}$, 
we have the fractional electric charge violating the Dirac quantization condition
on the axionic domain wall.
Therefore, such a configuration cannot be realized as a physical state.
Note that
the 0-form symmetry generator is 
reduced to the invertible symmetry generator
for $p_\phi = 1$.
In this case, 
the induced charge manifestly satisfies the Dirac quantization 
condition, 
$q_\phi q_{a\mathrm{M}} \in \bb{Z}$.

We remark that the non-invertible symmetry transformation on 
the 't Hooft loop is consistent 
with the monopole solution to the axionic domain wall problem,
where the axionic domain wall with $\bb{Z}_{k_\phi}$
charge is not stable, but only $\bb{Z}_{K_\phi}$ 
can be stable~\cite{Sato:2018nqy}. 
In other words, the axionic domain wall 
that is identified as the invertible 0-form symmetry generator 
can be stable.
The linked configuration with $k_\phi =1$ discussed in 
Refs.~\cite{Kogan:1992bq,Kogan:1993yw}
trivially satisfies our selection rule, i.e., 
the Dirac quantization condition 
because 
the (non-)invertible symmetry generator with $k_\phi =1$ is trivial.

\subsection{Selection rule for magnetic strings with axionic strings\label{axmag}}

Next, we show that the linked configurations 
of the magnetic and axionic strings are prohibited 
depending on the charge of the axionic strings.
To see this, we 
consider the non-invertible 1-form transformation 
on the axionic string with the linked configuration~\cite{Choi:2022fgx}.

\begin{figure}[t]
 \begin{center}
  \ig[height=10em]{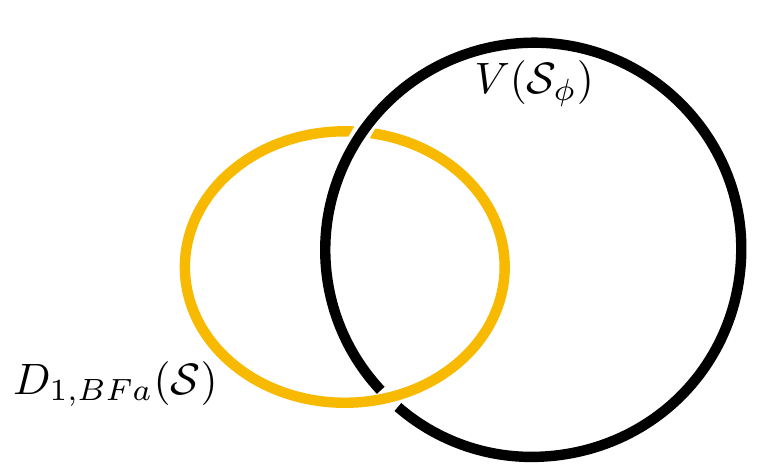}
 \end{center}
\caption{\label{mag-ax-strings}
Linked configuration of the magnetic string and axionic string
 in \er{240804.1618}.
This is a time slice of the worldsheet of the 
magnetic string (yellow circle) and axionic string (black circle).
The parameters of the objects are abbreviated.
In this figure, there should be another pair of the linked strings
with opposite orientations, but we omit it for simplicity.
}
\end{figure}

We consider the correlation function
$ \vevs{ D_{1, BFa} (e^{\fr{2\pi i n_1 }{k_a}} , {\mathcal{S}}) V (q_{\phi \mathrm{M}}, {\mathcal{S}}_\phi)} $, 
where 
${\mathcal{S}}$
and ${\mathcal{S}}_\phi$
are linked with each other 
at any time slice.
For example, 
we assume that 
both ${\mathcal{S}}$ and ${\mathcal{S}}_\phi$ have the topology of $S^1 \times S^1$,
and are 
extended 
along the temporal direction denoted as 
$\mathcal{S} = S^1_{a, \mathrm{time}} \times 
S^1_{a, \mathrm{space}}
$
and 
$\mathcal{S}_\phi  = S^1_{\phi , \mathrm{time}} \times 
S^1_{\phi , \mathrm{space}}
$ (see Fig.~\ref{mag-ax-strings}).
In this case, 
the correlation function can 
be evaluated as 
\begin{equation}
\vevs{ D_{1, BFa} (e^{\fr{2\pi i n_1 }{k_a}} , {\mathcal{S}}) V (q_{\phi \mathrm{M}}, {\mathcal{S}}_\phi)}
=  \vevs{ V (q_{\phi \mathrm{M}}, {\mathcal{S}}_\phi)
e^{  - \fr{ i q_a' q_{\phi \mathrm{M}}}{p_a'}\int_{S^1_{a , \mathrm{time}}} a}} 
\delta^{\mathrm{mod}\, p_a}_{q_{\phi \mathrm{M}},0}.
\label{240804.1618}
\end{equation}

The action of
the non-invertible 1-form symmetry generator 
on the axionic string 
means that the transition amplitude of the magnetic string linked with 
the axionic vortex is zero
unless $q_{\phi \mathrm{M}}
\in p'_a \bb{Z}$.
If $q_{\phi \mathrm{M}}
\in p'_a \bb{Z}$ holds, 
the Wilson loop
with a properly quantized 
charge $ - q'_a q_{\phi \mathrm{M}}/ p'_a$
is induced.
Physically, the configuration is cannot be realized 
even if each object is protected by topological charges.

\begin{figure}[t]
 \ig[width=35em]{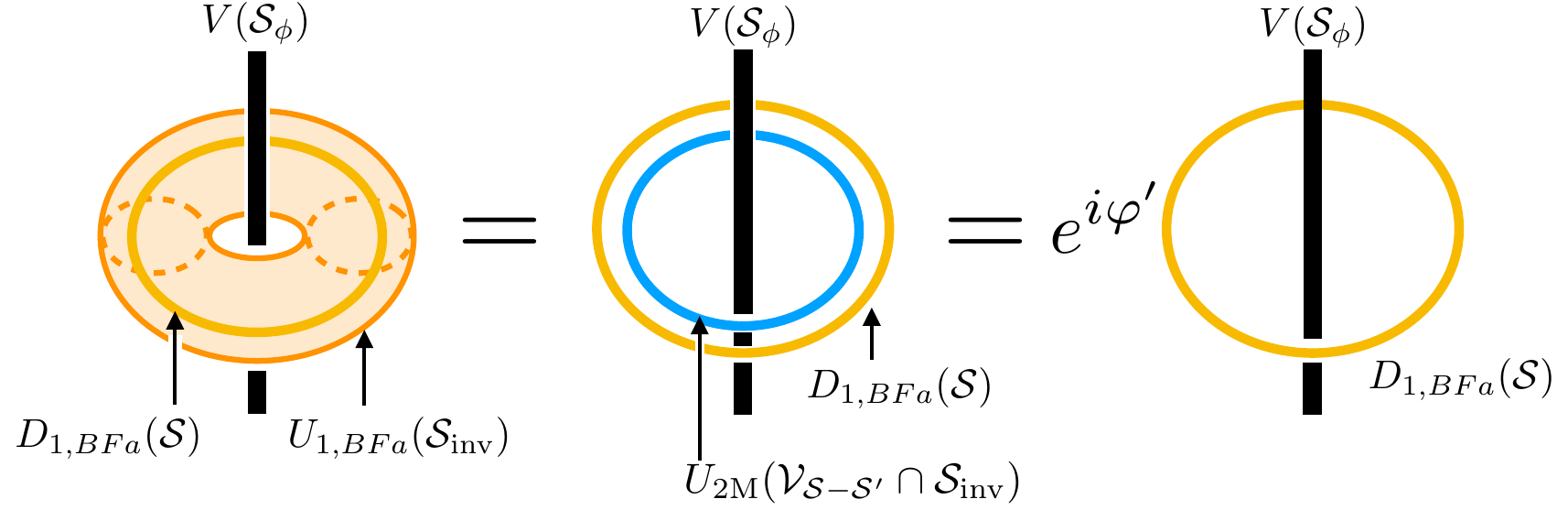}
\caption{\label{inv1-noninv1}
The correlation function of a worldsheet of the axionic string, 
 non-invertible 1-form symmetry generator, 
and invertible electric 1-form symmetry generator 
in \ers{240716.0445}--\eqref{240716.0509}.
In the left panel, 
we add the invertible electric 1-form symmetry generator expressed by 
the instantaneous yellow torus to the configuration in 
Fig.~\ref{mag-ax-strings}.
The torus encloses the magnetic string so that the invertible 
electric 1-form symmetry generator captures 
the electric flux emitted from the magnetic string.
In the middle panel, we have the induced 2-form symmetry generator 
when the non-invertible 1-form symmetry generator goes through 
the invertible electric 1-form symmetry generator.
In the right panel we have the phase 
$e^{i\varphi '} 
= e^{ - \fr{2\pi i m'_1}{K_a} \cdot \fr{N n_1}{k_a} \cdot q_{\phi \mathrm{M}}} $
in \er{240716.0509} by the action of the 2-form symmetry generator 
to the axionic string.
}
\end{figure}

As in the discussion on the 't Hooft loop and the axionic domain wall, 
we can understand the vanishing of the correlation function as 
a selection rule using the invertible electric 1-form global symmetry.
To see this, we try to capture the electric flux emitting from 
the magnetic string, i.e., the non-invertible 1-form symmetry generator
using an invertible electric 1-form symmetry generator (Fig.~\ref{inv1-noninv1}),
\begin{equation}
  \vevs{ U_{1,BF a} (e^{\fr{2\pi i m'_1}{K_a}} , {\mathcal{S}}_\mathrm{inv})
D_{1, BFa} (e^{\fr{2\pi i n_1 }{k_a}} , {\mathcal{S}}) V (q_{\phi \mathrm{M}}, {\mathcal{S}}_\phi)} .
\label{240716.0445}
\end{equation}
As in ordinary electromagnetism,
the invertible electric 1-form symmetry generator captures 
the electric flux emitted from the non-invertible 1-form symmetry generator.
We then move the non-invertible 1-form symmetry generator 
${\mathcal{S}} \to \mathcal{ S'}$ 
so that the invertible electric 1-form symmetry generator can shrink.
By this deformation, the non-invertible 1-form symmetry generator 
acts on the invertible electric 1-form symmetry generator 
as an invertible translation 
$a \to a - \fr{2 \pi n_1 }{k_a} \delta_1 
({\mathcal{V}}_{\mathcal{ S}- {\mathcal{S'}}}) $,
where ${\mathcal{V}}_{\mathcal{ S}- {\mathcal{S}}'}$ is a 3-dimensional space 
with $\der {\mathcal{V}}_{\mathcal{ S}- {\mathcal{S'}}} = \mathcal{ S} \cup \bar{\mathcal{S}}'$.
By the action, the invertible magnetic 2-form symmetry generator is 
induced, 
\begin{equation}
\begin{split}
&
  \vevs{U_{1, BF a} 
(e^{\fr{2\pi i m'_1}{K_a}}, {\mathcal{S}}_\mathrm{inv})
D_{1, BF a} (e^{\fr{2\pi i n_1 }{k_a}}, {\mathcal{S}}) V(q_{\phi \mathrm{M}}, {\mathcal{S}}_\phi)}
\\
&
= 
  \vevs{U_\mathrm{2 M} 
(e^{ - \fr{2\pi i m'_1}{K_a}\cdot \fr{N n_1}{k_a}},
{\mathcal{V}}_{\mathcal{ S} - {\mathcal{S}}'} \cap {\mathcal{S}}_\mathrm{inv} )
D_{1, BF a} (e^{\fr{2\pi i n_1 }{k_a}}, {\mathcal{S}}') V(q_{\phi \mathrm{M}}, {\mathcal{S}}_\phi)
} 
\end{split}
\end{equation}
Here, ${\mathcal{V}}_{\mathcal{ S} - {\mathcal{S}}'} \cap {\mathcal{S}}_\mathrm{inv} 
$
is a closed 1-dimensional subspace
by the choice of $ {\mathcal{V}}_{\mathcal{ S} - {\mathcal{S}}'}$.
Since the magnetic 2-form symmetry generator only acts on the 
worldsheet of the axionic string
in the correlation function, 
we arrive at the relation
after moving $\mathcal{ S'}$ to ${\mathcal{S}}$ again,
\begin{equation}
\begin{split}
&
  \vevs{U_{1, BF a} 
(e^{\fr{2\pi i m'_1}{K_a}}, {\mathcal{V}}_{\mathcal{ S} - {\mathcal{S}}'} \cap {\mathcal{S}}_\mathrm{inv} 
)
D_{1, BF a} (e^{\fr{2\pi i n_1 }{k_a}}, {\mathcal{S}}') V(q_{\phi \mathrm{M}}, {\mathcal{S}}_\phi)}
\\
&
= 
e^{ - \fr{2\pi i m'_1}{K_a} \cdot \fr{N n_1}{k_a} \cdot q_{\phi \mathrm{M}}}
  \vevs{
D_{1, BF a} (e^{\fr{2\pi i n_1 }{k_a}}, {\mathcal{S}}) V(q_{\phi \mathrm{M}}, {\mathcal{S}}_\phi)} .
\end{split}
\label{240716.0509}
\end{equation}
However, we have the same problem in the correlation function
as in the case of the axionic domain wall.
Since the invertible electric 1-form symmetry generator is parameterized by 
$e^{\fr{- 2\pi i m'_1}{K_a}} \in \bb{Z}_{K_a}$, 
the correlation function should be invariant under
$m'_1 \to m'_1 + K_a$.
The invariance requires that 
$e^{-2\pi i  \cdot \fr{N n_1}{k_a} \cdot q_{\phi \mathrm{M}}} = 
e^{ - 2\pi i \fr{q'_a}{p'_a} q_{\phi \mathrm{M}}} =1$.
This is only satisfied for $q_{\phi\mathrm{M}}  \in p'_a \bb{Z}$.
Otherwise, the correlation function becomes ambiguous, 
i.e., the $\bb{Z}_{p'_a}$ phase cannot be determined.
To avoid this ambiguity, the correlation function
$ \vevs{
D_{1, BF a} (e^{\fr{2\pi i n_1 }{k_a}}, {\mathcal{S}}) V(q_{\phi \mathrm{M}}, {\mathcal{S}}_\phi)} $ 
should vanish,
\begin{equation}
 \vevs{
D_{1, BF a} (e^{\fr{2\pi i n_1 }{k_a}}, {\mathcal{S}}) V(q_{\phi \mathrm{M}}, {\mathcal{S}}_\phi)}
 = 0 ,\qtq{for} q_{\phi \mathrm{M}} \not\in  p'_a \bb{Z}.
\end{equation}

Again, we can relate the non-invertible action to the Dirac quantization 
condition.
We consider the induced charge 
$ \fr{N}{4\pi^2} \int_\mathcal{ V_{S_\mathrm{inv}}} \bs{\na} \phi \cdot \bs{B} dV $ 
on the magnetic string,
where $\mathcal{ V_{S_\mathrm{inv}}}$ is a 3-dimensional subspace 
with $\der \mathcal{ V_{S_\mathrm{inv}}} = {\mathcal{S}}_\mathrm{inv}$.
By substituting the configuration of the axionic string and the 
magnetic string, we have 
$ \fr{N}{4\pi^2} \int_\mathcal{ V_{S_\mathrm{inv}}} \bs{\na} \phi \cdot \bs{B} dV =   - \fr{N n_1 q_{\phi \mathrm{M}}}{k_a} =  - \fr{q'_a q_{\phi \mathrm{M}} }{p'_a}$.
Therefore, if $q_{\phi \mathrm{M}} \not\in p'_a \bb{Z}$, 
the charge on the magnetic string 
violates the Dirac quantization condition, 
and such a state is prohibited.
The non-invertible action on the axionic string represents the restriction 
on the physical state by the Dirac quantization.

\subsection{Implications to solitons with finite widths}

So far, we have considered the low-energy limit of the systems
where the width of the axionic domain walls or the magnetic strings
can be 
neglected.
Meanwhile, the topological solitons can have finite widths
in a finite energy scale.
Therefore, it will be a non-trivial question whether we can apply 
our discussion is based on the low-energy limit 
to the finite energy scale.

We can show that the configuration of the topological solitons 
violating the Dirac quantization is prohibited by contradiction
even for the finite energy scale.
For example, we consider the linked configuration of
 the worldsheet of the axionic and magnetic strings.
We assume that there is no phase transition between 
the low-energy limit 
and the finite 
energy region where the magnetic strings 
with a finite width exist.
We consider a linked configuration of the axionic
and the magnetic strings.
We can evaluate the induced electric charge on the magnetic strings by 
the volume integral of the whole space other than the singularity 
on which the core of the axionic string exists,
$\fr{N}{4\pi^2} \int_{\bb{R}^3 \backslash {\mathcal{S}}_\phi} 
d^3 \bs{x} \bs{\na} \phi \cdot \bs{B}$.
We then take the low-energy limit where the magnetic string is localized 
on a 2-dimensional surface in the spacetime.
Since the electric charge is quantized, 
it cannot be changed under the continuous deformation of the energy scale.
Since there is no phase transition, we can avoid new dynamical fields
which can carry electric charges.
Therefore, the electric charge induced on the magnetic string
for the finite energy scale 
should match that for the low-energy limit.
Since the magnetic string is localized, we have
\begin{equation}
 \fr{N}{4\pi^2} \int_{\bb{R}^3 \backslash {\mathcal{S}}_\phi} 
d^3 \bs{x} \bs{\na} \phi \cdot \bs{B}
= 
\fr{N}{4\pi^2} \int_\mathcal{ V_{S_\mathrm{inv}}} \bs{\na} \phi \cdot \bs{B}d^3 \bs{x}.
\end{equation}
If the induced electric charge violates the Dirac quantization, 
the non-invertible symmetry shows that it cannot exist in the 
low-energy limit.
Therefore, there is a contradiction between 
the existence of the linked configuration violating 
the Dirac quantization in the finite energy scale 
and the non-invertible 1-form symmetry transformation 
on the axionic string.

We remark that the same argument holds for 
the axionic domain wall with a finite width 
enclosing the magnetic monopole.
If the configuration violates the Dirac quantization, it cannot be 
realized because it contradicts the non-invertible 0-form symmetry 
transformation after taking the low-energy limit.

\section{\label{sec:summary}Summary and discussion} 

In this paper, we have discussed the selection rules of the topological solitons in the presence of the magnetic monopoles and the axionic strings
in the $(3+1)$-dimensional axion electrodynamics 
 from the viewpoint of the non-invertible global symmetries. 
After constructing the non-invertible symmetry generators in the massive axion and the massive photon phases, we have identified the non-invertible 0-form symmetry generator as a worldvolume of an axionic domain wall, and the non-invertible 1-form symmetry generator as a worldsheet of a magnetic string, respectively.
At the expense of the invertibility, the non-invertible symmetry generators correspond to the topological solitons with any charge.

By the vanishing of the correlation function involving the non-invertible 0-form symmetry generator and the 't Hooft loop, we have shown that the configuration of the axionic domain wall enclosing the magnetic monopole 
with the condition 
$q_{a \mathrm{M}} \not\in p_\phi \bb{Z}$ in \er{240917.1539}
is prohibited.
In the massive photon phase,
we have further shown 
that the linked configuration of the axionic and magnetic strings 
the condition 
$q_{\phi \mathrm{M}} \not\in p'_a \bb{Z}$ in \er{240804.1618}
is prohibited
using the vanishing of the correlation function between the non-invertible 1-form symmetry generator and the worldsheet of the axion string.

We have argued that the above vanishing correlation functions can be understood as selection rules for the topological solitons.
Physically, the selection rule is due to the violation of the Dirac quantization of the electric charge induced on the axionic domain wall or the magnetic string.
The selection rules have been expressed in terms of the invertible electric  1-form symmetry generator, which captures the electric flux emitted from the topological solitons.
Finally, we have shown that our selection rules in the low-energy limit can be promoted to those in finite energy scales where solitons have finite widths. 

There are several avenues for future work.
We can apply our constraints to cosmology with specific models which can have a configuration of 
the pair of axionic domain walls and monopoles~\cite{Sato:2018nqy}, or the pair of axionic and magnetic strings~\cite{Eto:2024hwn}.
While the magnetic monopoles and the axionic strings are treated as singularities in this paper, 
we can apply our discussion to systems that include 't~Hooft-Polyakov monopoles or axionic vortices with finite widths, and directly discuss our claim in section~\ref{conf}.
It would be interesting to extend our discussion to other systems with non-invertible symmetry generators that can be regarded as topological solitons, and to explore similar constraints on 
the configurations of solitons.

\section*{Acknowledgments}

RY thanks Masaki Yamada for helpful discussions
during the Yukawa Institute for Theoretical Physics workshop YITP-W-24-09 on ``Progress in Particle Physics 2024.'' 
This work is supported in part by 
 JSPS KAKENHI [Grant Numbers JP22H01221 (MN), JP21K13928 (RY), JP21H01084 and JP24H00975 (YH)] 
and the WPI program ``Sustainability with Knotted Chiral Meta Matter 
(SKCM$^2$)'' at Hiroshima University (MN).
 
\appendix

\section{\label{appendix}Derivations of non-invertible symmetry transformations}

Here, we review the derivations of the non-invertible 0- 
and 1-form symmetry transformations used in \ers{240917.1539} 
and \eqref{240804.1618}, respectively.

\subsection{\label{0noninv}Non-invertible 0-form symmetry transformation on 't Hooft loop}

We here review the derivation of the correlation function
$    \vevs{
    D_{0, BF \phi }(e^{\fr{2 \pi i n_0 }{ k_\phi}} ,S^2 \times S^1)
    T(q_{a\mathrm{M}}, S^1) 
    }$
in \er{240917.1539} based on Refs.~\cite{Choi:2022jqy,Cordova:2022ieu}.
Since $ S^2 \times S^1$ of the non-invertible 0-form symmetry generator
encloses the 't Hooft loop $T(q_{a\mathrm{M}}, S^1) $,
the magnetic flux $da$ should satisfy 
the quantization condition on $S^2$, 
\begin{equation}
    \int_{S^2} da = 
     \int_{S^2} da_{\rm S} = 2\pi q_{a \mathrm{M}},
\end{equation}
where $a_{\rm S}$ is the singular part of the photon in the presence 
of the 't Hooft loop in \er{241016.0441}.
Substituting the configuration into 
the 0-form symmetry generator in \er{240314.1946}, we have
\begin{equation}
\fr{1}{2\pi }  
\int_{S^2 \times S^1} u_i^\phi \wed da 
= 
\fr{1}{2\pi }  
\int_{S^2 \times S^1} u_i^\phi \wed da_\mathrm{R} 
+
 q_{a {\rm M}}
\int_{S^1} u_i^\phi 
.
\end{equation}
Here, $a_{\rm R}$ is the regular part of the photon satisfying 
$dda_{\rm R}=  0$.
Therefore, the insertion of the 't Hooft loop 
corresponds to the insertion of the Wilson loop of $u_i^\phi$ with the charge $q_{a \mathrm{M}}$.
Now, we evaluate the partition function 
associated with the 0-form symmetry generator.
For the regular part, we can safely absorb $a_\mathrm{R}$
by $u^i_\phi - \fr{1}{p_\phi} a_\mathrm{R} \to u_i^\phi $ 
without violating the Dirac quantization of $ u_i^\phi$, 
and 
we have 
$e^{ - i \fr{N n_0}{4\pi k_\phi }\int_{S^2 \times S^1} a_\mathrm{R} \wed d a_\mathrm{R} -\fr{i q_{a \mathrm{M}} q_\phi}{p_\phi} \int_{S^1} a_\mathrm{R} } $.
Since the term 
$e^{ - i n_0 \int_{S^2 \times S^1} c - i \fr{N n_0}{4\pi k_\phi }\int_{S^2 \times S^1} a_\mathrm{R} \wed d a_{\rm R}}$
can be absorbed into the action by the redefinition of $\phi$ as in the invertible 0-form symmetry, 
we have $e^{-\fr{i q_{a \mathrm{M}} q_\phi}{p_\phi} \int_{S^1} a_\mathrm{R}}$
in the correlation function.
For the part $ q_{a {\rm M}}
\int_{S^1} u_i^\phi $,
we evaluate the partition function,
which is the same as the expectation value of the Wilson loop 
in the level-$p_\phi$ $U(1)$ Chern-Simons theory, which shows that~\cite{Witten:1988hf}
\begin{equation}
\mathcal{ N}_0 \int \mathcal{ D}u^\phi_1... \mathcal{ D} u^\phi_{ q_\phi}
\exp\( i
\sum_{i = 1}^{q_\phi}\int_{S^2 \times S^1} 
\(
  \fr{p_\phi}{4\pi}  u^\phi_i \wed du^\phi_i
 - q_{a \mathrm{M}}\int_{S^1} 
u_i^\phi
\) \)
= \delta_{q_{a \mathrm{M}}, 0}^{\mathrm{mod}\, p_\phi} ,
\end{equation}
and we have the correlation function in \er{240917.1539}.
Note that the normalization factor $\mathcal{ N}_0$
is fixed so that the right-hand side of this equation is 
$\delta_{q_{a \mathrm{M}}, 0}^{\mathrm{mod}\, p_\phi}$.
Therefore, the correlation function of the 0-form symmetry generator and 
't~Hooft loop with $q_{a \mathrm{M}}  \not\in  p_\phi  \bb{Z}$ vanishes. 

\subsection{Non-invertible 1-form symmetry transformation on axionic string}

Next, we review 
non-invertible 1-form symmetry transformation on 
an axionic string~\cite{Choi:2022fgx}
represented by the correlation function $\vevs{ D_{1, BFa} (e^{\fr{2\pi i n_1 }{k_a}} , {\mathcal{S}}) V (q_{\phi \mathrm{M}}, {\mathcal{S}}_\phi)}$
in \er{240804.1618}.
To do this, we decompose the axion field into 
the regular part and the singular part representing the fluctuation and the axionic string, respectively.
In the presence of the axionic string, 
the axion has a winding number along  
a circle $S^1$
around 
$\mathcal{S}_\phi$,
\begin{equation}
    \int_{S^1_{a , \mathrm{space}}} d\phi =  2 \pi q_{\phi \mathrm{M}},
\end{equation}
and therefore, $\phi$ becomes a multivalued function.
We can decompose the axion into  the regular part
and the singular part,
\begin{equation}
    d \phi
    = d \phi_{\mathrm{R}}
    + d\phi_{\rm S},
\end{equation}
where
\begin{equation}
    \int_{S^1_{a , \mathrm{space}}}
     d \phi
     = 
     \int_{S^1_{a , \mathrm{space}}}
     d \phi_{\rm S}
    =2 \pi q_{\phi{\rm M}}.
\end{equation}
We then substitute the configuration to 
$  \fr{1}{2 \pi }
    \int_{\mathcal{S}} u^a_i \wed d\phi$
    in the partition function 
$Z_1 [p'_a, q'_a, S^1_{a, \mathrm{time}} \times 
S^1_{a, \mathrm{space}}, da, d\phi ] $ in \er{241008.0658},
and we have
\begin{equation}
    \fr{1}{2 \pi }
    \int_{S^1_{a, \mathrm{time}} \times 
S^1_{a, \mathrm{space}}} u^a_i \wed d\phi
    =
     \fr{1}{2 \pi }
    \int_{S^1_{a, \mathrm{time}} \times 
S^1_{a, \mathrm{space}}}u^a_i \wed d\phi_{\mathrm{R}}
    + q_{\phi \mathrm{M}}
    \int_{S^1_{a , \mathrm{time}}} u^a_i .
\end{equation}
Therefore, the axionic string 
induces the Wilson loop for 
$u^a_i$ in the partition function with the charge $q_{\phi \mathrm{M}}$.
For the regular part, we 
can safely absorb it 
to $\chi_i^a$ by 
the redefinition 
$\chi_i^a - \fr{1}{p'_a} \phi_{\mathrm{R}}
 \to \chi_i^a$
 without violating the 
 flux quantization condition 
 for $\chi_i^a$.
Similarly, we can absorb 
the gauge field $a$
by $u_i^a - \fr{1}{p'_a} a \to u_i^a $
because of the absence of 
the magnetic monopole, 
$\int_{\mathcal{S}} da =0$.
By the redefinitions, 
we have 
\begin{equation}
\begin{split}
&
\int \mathcal{ D} \chi_i^a 
\mathcal{ D} u^a_i
\exp \(\fr{i}{2\pi} 
\sum_{i =1}^{q'_a}\int_{S^1_{a, \mathrm{time}} \times 
S^1_{a, \mathrm{space}}} (p'_a \chi^a_i du^a_i - 
\chi^a_i da -  u_i^a \wed d\phi) \)
\\
&
= 
\exp 
\(- \fr{iq'_a}{2 \pi p'_a}
\int_{S^1_{a, \mathrm{time}} \times 
S^1_{a, \mathrm{space}}} 
\phi_{\mathrm{R}} da 
- \fr{ i q'_a q_{a \mathrm{M}}}{p_a'}\int_{S^1_{a , \mathrm{time}}} a
\)
\\
&
\qquad
\times 
\int \mathcal{ D} \chi_i^a 
\mathcal{ D} u^a_i
\exp \(\fr{i}{2\pi} 
\sum_{i = 1}^{q'_a}
\int_{S^1_{a, \mathrm{time}} \times 
S^1_{a, \mathrm{space}}} 
p'_a \chi^a_i du^a_i  
- i q_{\phi \mathrm{M}}
\sum_{i = 1}^{q'_a}
\int_{S^1_{a , \mathrm{time}}} u^a_i \).
\end{split}
\end{equation}
The path integral of 
$u_i^a$ gives us the 
constraint by the delta function (see, e.g.,~Ref.~\cite{Thorngren:2018ziu})
\begin{equation}
    \int 
\mathcal{ D} u^a_i
\exp \(\fr{i}{2\pi} 
\sum_{i =1}^{q'_a}
\int_{S^1_{a, \mathrm{time}} \times 
S^1_{a, \mathrm{space}}} 
p'_a \chi^a_i du^a_i  
- i q_{\phi \mathrm{M}}
\sum_{i =1}^{q'_a}
\int_{S^1_{a , \mathrm{time}}} u^a_i \)
 \propto
\delta_{q_{\phi \mathrm{M}}, 0}^{\mathrm{mod}\, p'_a} .
\end{equation}
By choosing the normalization factor ${\cal N}_1$ appropriately, 
the path integral of 
$\chi^a_i$ becomes
\begin{equation}
\begin{split}
&
{\cal N}_1 
\int \mathcal{ D} \chi_i^a 
\mathcal{ D} u^a_i
\exp \(\fr{i}{2\pi} 
\sum_{i =1}^{q'_a}
\int_{S^1_{a, \mathrm{time}} \times 
S^1_{a, \mathrm{space}}} (p'_a \chi^a_i du^a_i - 
\chi^a_i da -  u_i^a \wed d\phi) \)
\\
&
=
\exp 
\(- \fr{iq'_a}{2 \pi p'_a}
\int_{S^1_{a, \mathrm{time}} \times 
S^1_{a, \mathrm{space}}} 
\phi_{\mathrm{R}} da 
- \fr{ i q'_a q_{ \phi \mathrm{M}}}{p_a'}\int_{S^1_{a , \mathrm{time}}} a
\)\delta^{\mathrm{mod}\, p'_a}_{q_{\phi \mathrm{M}},0 }.
\end{split}
\end{equation}
Since 
$\fr{iq'_a}{2 \pi p'_a}
\int_{S^1_{a, \mathrm{time}} \times 
S^1_{a, \mathrm{space}}} 
\phi_{\mathrm{R}} da $ 
can be absorbed into the action by the redefinition 
$a + \fr{2\pi n_1}{k_a}\delta_1 ({\cal V_S}) \to a$
with ${\cal S} = S^1_{a, \mathrm{time}} \times 
S^1_{a, \mathrm{space}}$, 
 we obtain \er{240804.1618}.

\providecommand{\href}[2]{#2}
\end{document}